\documentclass{article}
\usepackage{amsmath}

\input{tcilatex}
\begin{document}

\author{Giuseppe Castagnoli \\
Elsag Bailey ICT Division and Quantum Information Laboratory\thanks{%
retired} }
\title{An exact relation between number of black box computations required
to solve an oracle problem quantumly and quantum retrocausality}
\maketitle

\begin{abstract}
We investigate the reason for the quantum computational speedup -- quantum
algorithms requiring fewer computation steps than their classical
counterparts. We extend the representation of the quantum algorithm to the
process of setting the problem -- of choosing the function computed by the
black box. The initial measurement selects a setting at random, Bob (the
problem setter) unitarily changes it into the desired one. This
representation is to Bob and any external observer, it cannot be to Alice
(the problem solver). It would tell her the function computed by the black
box, which to her should be hidden inside it. We resort to the observer
dependent quantum states of relational quantum mechanics. To Alice, the
projection of the quantum state due to the initial measurement is retarded
at the end of her problem solving action. To her, the algorithm input state
remains one of complete ignorance of the setting. By black box computations,
she unitarily sends it into the output state that, for each possible
setting, encodes the corresponding solution, acquired by the final
measurement. We show that there can be a quantum feedback from final to
initial measurement. We can ascribe to the final measurement the selection
of any part -- say the $\mathcal{R}$-th part -- of the random outcome of the
initial measurement. This projects the input state to Alice on a state of
lower entropy where she knows a corresponding part of the \ problem setting.
The quantum algorithm turns out to be a sum over classical histories in each
of which Alice, knowing in advance one of the $\mathcal{R}$-th parts of the
setting, performs the black box computations still required to identify the
solution. Given an oracle problem and a value of $\mathcal{R}$, this
retrocausality model provides the number of black box computations required
to solve it quantumly. Conversely, given a known quantum algorithm, it
yields the value of $\mathcal{R}$ that explains its speed up. In the major
quantum algorithms, $\mathcal{R}$ is $\frac{1}{2}$ or slightly above it. $%
\mathcal{R}=\frac{1}{2}$ always yields the number of black box computations
required by an existing quantum algorithm\ and the order of magnitude of the
number required by optimal one.
\end{abstract}

\section{Foreword}

Consider the following problem. Bob, the problem setter, chooses one of the
four functions $f_{\mathbf{b}}\left( \mathbf{a}\right) $ whose tables are
given in array (\ref{deutsch}):%
\begin{equation}
\begin{tabular}{|l|l|l|l|l|}
\hline
$\mathbf{a}$ & $f_{00}\left( \mathbf{a}\right) $ & $f_{01}\left( \mathbf{a}%
\right) $ & $f_{10}\left( \mathbf{a}\right) $ & $f_{11}\left( \mathbf{a}%
\right) $ \\ \hline
$0$ & $0$ & $0$ & $1$ & $1$ \\ \hline
$1$ & $0$ & $1$ & $0$ & $1$ \\ \hline
\end{tabular}
\label{deutsch}
\end{equation}%
Then he gives Alice, the problem solver, a black box (\textit{oracle}) that,
given a value of the argument $\mathbf{a}$ in the input, produces the value
of $f_{\mathbf{b}}\left( \mathbf{a}\right) $ in the output. Alice does not
know which of the four functions is the one computed by the black box. She
is to determine whether the function is constant or balanced (ie with the
same number of zeros and ones) by performing function evaluations (\textit{%
oracle queries}). Classically, Alice must perform two function evaluations,
quantumly just one. We are speaking of the seminal quantum algorithm that
yields a \textit{quantum computational speedup}, devised by Deutsch $\left[ 1%
\right] $ in 1985.

Although there is a significant body of literature on the relationship
between speedup and other quantum features, such as \textit{quantum
entanglement} and \textit{discord} (see Section 3), no fundamental physical
explanation nor unified mathematical mechanism is known for the speedup.

As the present issue of Quanta is dedicated to Feynman and our subject is
quantum computation, we should like to remember Feynman's pioneering
contribution to the development of this new branch of science.

We do this by recalling the seminal works that gave rise to the discipline
of quantum computation. In 1969, Finkelstein $\left[ 2\right] $ noted that
computation should be possible in the quantum framework and introduced the
notion of quantum unit of information, namely quantum bit or \textit{qubit}.
In 1982 Feynman $\left[ 3\right] $ pointed out the essential difference
between quantum and classical computation, showing that the simulation of a
quantum process on a classical computer has to involve in general an amount
of time$\times $physical resources exponentially higher than that involved
in the quantum process itself. This was the origin of the notion of quantum
computational efficiency, now usually called speedup. Always in $\left[ 3%
\right] $, Feynman introduced the universal quantum simulator, a lattice of
spin systems with freely specifiable nearest neighbor interactions that can
be considered the first theoretical model of a quantum computer. The
development of the notion of reversible classical computation was parallel.
Bennett $\left[ 4\right] $ showed in 1982 that classical computation can be
ideally reversible in the limit of zero speed. His work was in the wake of
the 1961 \textit{Landauer's principle} $\left[ 5\right] $\ that quantifies
the generation of heat necessarily consequent to the erasure of information.
Still in 1982, Fredkin and Toffoli $\left[ 6\right] $ developed the first
theoretical model of logically reversible classical computation, in fact
based on the well known Fredkin and Toffoli gates. Independently of Deutsch,
Feynman produced in 1985 the quantum version of this algorithmic form of
reversible computation, published in the following year $\left[ 7\right] $.
The seminal idea of it was already present in $\left[ 3\right] $, in fact
with reference to Fredkin\&Toffoli 1982 work. We already cited the 1985
Deutsch's paper with the first example of a quantum algorithm that requires
fewer function evaluations than classically possible.

With this, the full fledged notion of quantum computation was born thanks to
the insights of very few individuals. As it might happen with revolutionary
science, the scientific community at large has been initially slow in
expressing an interest for the new discipline. We had the fortune of
contributing, with Mario Rasetti, to the organization of the first
international meetings on quantum communication and computation held in
Turin in the years 1992$\div $1998. They were the Elsag Bailey (an Italian
ICT company)-ISI (Institute for Scientific Interchange) workshops. We
believe that those annual workshops have been instrumental to propagating
the new discipline of quantum information throughout the scientific
community. All the fathers of the discipline, with the sad exception of
Feynman, the theoretical and experimental physicists and computer scientists
responsible for the major developments of those years attended the workshops
in question. The group pictures of reference $\left[ 8\right] $ show the
explosion of interest for the new science in the years 1993 through 1997.

\section{Introduction}

The usual physical representation of quantum algorithms is limited to the
process of solving the problem. We extend it to the process of setting the
problem, namely of choosing the function $f_{\mathbf{b}}\left( \mathbf{a}%
\right) $ out of the set of functions. This amounts to choosing the function
suffix $\mathbf{b}$, which we call \textit{the} \textit{problem setting},
out of the set of the possible problem settings $\sigma _{B}\equiv \left\{
00,01,10,11\right\} $ -- we use Deutsch algorithm as an example.

For reasons that will soon become clear, we assume that the initial state of
the quantum register $B$ that contains the setting is a mixture of all the
possible settings. Its density operator is thus $\rho _{B}=\frac{1}{4}\left(
\left\vert 00\right\rangle \left\langle 00\right\vert _{B}+\left\vert
01\right\rangle \left\langle 01\right\vert _{B}+\left\vert 10\right\rangle
\left\langle 10\right\vert _{B}+\left\vert 11\right\rangle \left\langle
11\right\vert _{B}\right) $ -- a few mathematical representations make
things more clear.

At time $t_{0}$, Bob measures the content of register $B$, obtaining a
setting at random, say $\mathbf{b}=10$. The state of register $B$\ is
consequently projected on $\left\vert 10\right\rangle \left\langle
10\right\vert _{B}$. Assume that Bob wants $\mathbf{b}=01$. He unitarily
transforms this state into $\left\vert 01\right\rangle \left\langle
01\right\vert _{B}$, at time $t_{1}$.

Register $A$, meant to contain the argument of the function to be computed
by the black box and eventually the solution of the problem, at time $t_{1}$
is in any sharp state, say $\left\vert 0\right\rangle \left\langle
0\right\vert _{A}$. The input state of the quantum algorithm at time $t_{1}$
is thus $\left\vert 01\right\rangle \left\langle 01\right\vert _{B}\otimes
\left\vert 0\right\rangle \left\langle 0\right\vert _{A}$.

Alice, with one function evaluation preceded and followed by suitable
transformations, unitarily transforms it into the output state $\left\vert
01\right\rangle \left\langle 01\right\vert _{B}\otimes \left\vert
1\right\rangle \left\langle 1\right\vert _{A}$, at time $t_{2}$. The
solution of the problem, $1$ when the function is balanced as in the present
case and $0$ when it is constant, is in register $A$. Alice acquires the
solution by measuring the content of $A$.

We note that this extended representation immediately calls for another
extension, this time concerning the actors (observers) on the stage. We have
to resort to the relational quantum mechanics of Rovelli $\left[ 9\right] $,
where quantum states are observer dependent. A quantum state can be sharp to
an observer and a quantum superposition, or a mixture, to another one. The
present representation is with respect to Bob, the problem setter, and any
other observer who does not act on the problem solving process. It cannot be
with respect to Alice, the problem solver. The input state of the quantum
algorithm $\left\vert 01\right\rangle \left\langle 01\right\vert _{B}\otimes
\left\vert 0\right\rangle \left\langle 0\right\vert _{A}$ would of course
tell us that the content of register $B$ is $01$, namely that the function
chosen by Bob is $f_{01}\left( \mathbf{a}\right) $. Since Alice is an
observer, we assume that the state in question would tell her the same.

Throughout this work, we take for granted the legitimacy of the assumption
that Alice (or Bob, or the external observer), although an abstract entity,
knows what we would know in her (his) place. We assume that this is a
legitimate way to take the sizes to a quantum process that necessarily
involves the notion of observer.

In the present case, Alice would know that the function is balanced without
performing any function evaluation. Of course the suffix of the function
should be hidden to Alice -- to her it is inside the black box.

We physically represent this concealment by retarding the projection of the
quantum state induced by the initial Bob's measurement at the end of the
unitary part of Alice's problem solving action. As well known, these
projections can be retarded or advanced at will along a unitary evolution
that respectively follows or precedes the measurement.

The input state of the quantum algorithm to Alice, immediately after the
preparation of the desired problem setting, remains thus $\frac{1}{4}\left(
\left\vert 00\right\rangle \left\langle 00\right\vert _{B}+\left\vert
01\right\rangle \left\langle 01\right\vert _{B}+...\right) \otimes
\left\vert 0\right\rangle \left\langle 0\right\vert _{A}$. In fact the
maximally mixed state of register $B$ remains unaltered under any unitary
transformation applying to it. The two bit entropy of this state represents
Alice's complete ignorance of Bob's choice.

The output state to Alice is $\frac{1}{4}(\left\vert 00\right\rangle
\left\langle 00\right\vert _{B}\otimes \left\vert 0\right\rangle
\left\langle 0\right\vert _{A}+$ $\left\vert 01\right\rangle \left\langle
01\right\vert _{B}\otimes \left\vert 1\right\rangle \left\langle
1\right\vert _{A}+...)$, namely still a mixture of all the possible problem
settings, each multiplied by the corresponding solution. Thus, also the
solution, considered in itself, is completely undetermined. Alice's final
measurement projects this state on the solution corresponding to the problem
setting chosen by Bob, namely on $\frac{1}{2}\left( \left\vert
01\right\rangle \left\langle 01\right\vert _{B}+\left\vert 10\right\rangle
\left\langle 10\right\vert _{B}\right) \otimes \left\vert 1\right\rangle
\left\langle 1\right\vert _{A}$,\ with probability one. In fact the solution
is unpredictable to Alice but is already $1$\ to any other observer.

Alice's final measurement also triggers the retarded projection induced by
the initial Bob's measurement, which cannot go past the unitary part of
Alice's action. This further projects the above state on $\left\vert
01\right\rangle \left\langle 01\right\vert _{B}\otimes \left\vert
1\right\rangle \left\langle 1\right\vert _{A}$, which tells Alice both the
problem setting and the solution. The two projections commute and should be
considered simultaneous.

We note that either the projection of the quantum state induced by the
initial Bob's measurement or that induced by the final Alice's measurement
zeroes the entropy of the solution, depending on which one is performed
first. This work is an exploration of the assumption that this zeroing
shares in a complementary and non-redundant way between initial and final
measurement.

We assume that the complete measurements behave in a contextual way -- each
would be sensitive to the other. We assume that they reduce (in all the
possible ways in quantum superposition as we will see) to partial
measurements such that, together, select whatever has been selected by the
complete measurements and, each by itself, reduce the entropy of the
solution in a complementary and non-redundant way. For Occam razor, we
should exclude any redundancy. This implies that the information provided by
either partial measurement is not provided by the other.

To reconstruct the selections performed by the complete measurements, we
should propagate forward in time, along the time-forward unitary
transformation, the projection of the quantum state due to the partial Bob's
measurement, until it selects part of the outcome of Alice's measurement.
Similarly, we should propagate backward in time, along the inverse of the
time-forward unitary transformation, the projection due to Alice's partial
measurement, until it selects part of the random outcome of Bob's
measurement.

We will see that everything boils down to ascribing to the final Alice's
measurement the selection of part of the random outcome of the initial Bob's
measurement, say the $\mathcal{R}$-th part of the information that specifies
it.

This \textit{quantum feedback} leaves the input state of the quantum
algorithm to Bob and any external observer unaltered. It projects that to
Alice on a state of lower entropy where she knows the $\mathcal{R}$-th part
of the information that specifies the problem setting, before performing any
function evaluation. Alice uses this knowledge to solve the problem with
fewer function evaluations. In other words, sharing the selection of the
final measurement outcome between initial and final measurement explains the
speedup.

There are many ways of taking the part of a whole. The quantum algorithm
turns out to be a sum over classical histories in each of which Alice knows
in advance one of the possible $\mathcal{R}$-th parts of the problem setting
and performs the function evaluations still necessary to find the solution
of the problem.

We can speak of Alice's \textit{advanced knowledge} of part of the problem
setting because, in the absence of quantum retrocausality, Alice would know
the setting only with the final measurement. Moreover this knowledge comes
to her from her final reading of the solution.

Given an oracle problem and a value of $\mathcal{R}$, the present
retrocausal explanation of the speedup yields a number of function
evaluations required to solve the problem quantumly. Conversely, given a
known quantum algorithm, it yields the value of $\mathcal{R}$ that explains
its speedup.

We have compared this model with the major quantum algorithms discovered so
far. In all the quantum algorithms that solve the problem with a single
function evaluation, as that of Deutsch, we have $\mathcal{R}=\frac{1}{2}$.
This also applies to Grover quantum search algorithm for database size $4$,
Deutsch \& Jozsa algorithm, and the algorithms of Simon and the Abelian
hidden subgroup. The latter algorithm $\left[ 10\right] $ has unified about
ten historical algorithms, among which the famous Shor's factorization
algorithm. In Grover algorithm, when database size goes past $4$, first $%
\mathcal{R}$ goes slightly above $\frac{1}{2}$ then it goes back to $\frac{1%
}{2}$\ for database size tending to infinity.

In the corresponding sample of problems, $\mathcal{R}=\frac{1}{2}$ always
corresponds to an existing quantum algorithm and yields the order of
magnitude of the number of function evaluations required to solve the
problem in an optimal quantum way. If this held in general, we would have a
very powerful tool, the way of assessing the order of magnitude of the
number of function evaluations (oracle queries) required to solve a generic
oracle problem in an optimal quantum way.

\section{Positioning the work}

The present work is the further development of the approach $\left[ 11\div 13%
\right] $. We have further clarified the retrocausal explanation of the
speedup and developed a procedure for computing the number of function
evaluations required to solve a generic oracle problem with quantum
retrocausality $\mathcal{R}=\frac{1}{2}$.

The present explanation of the speedup is in line with the tenet of
time-symmetric quantum mechanics of Aharonov et al. $\left[ 14,15,16,17%
\right] $, which states that the complete description of the quantum process
between initial and final measurement requires knowledge not only of the
outcome of the initial measurement, also of that of the final one. This
naturally implies that the latter outcome has back in time implications on
the upstream process. As a matter of fact, the form of quantum
retrocausality utilized in the present work has been inspired by the work of
Dolev and Elitzur $\left[ 17\right] $ on the non-sequential behavior of the
wave function highlighted by partial measurement.

The work has points of contact with works of Morikoshi. In $\left[ 18\right] 
$, this author highlights the problem-solution symmetry of Grover's and the
phase estimation algorithms and notes it may be relevant for the explanation
of the speedup. In $\left[ 19\right] $, he shows that Grover algorithm
violates a temporal Bell inequality. There should be a connection between
this violation and the form of quantum retrocausality we are dealing with.

Besides $\left[ 11\div 13\right] $,\ we are not aware of literature relating
the speedup to quantum retrocausality. There are of course other approaches
to the problem of unifying the explanation of the speedup. Reference $\left[
20\right] $ (2001) shows that the presence of multipartite entanglement with
number of parties increasing unbounded with problem size is necessary for
achieving exponential speedup in pure state quantum computing. It also
conjectures that there could be exponential speedup in the absence of
entanglement in mixed state quantum computing.

The notion of quantum discord was introduced independently in $\left[ 21%
\right] $ and $\left[ 22\right] $. Discord is a measure of non-classical
correlations between two subsystems of a quantum system that are not
necessarily entangled -- it coincides with entanglement in pure state
quantum computing. It could be of high practical interest, since it shows
the possibility of achieving a speedup in mixed state quantum computing --
the realistic form of computation in the presence of noise. Reference $\left[
23\right] $ shows that, contrary to the topical thought at the time, quantum
states can be too entangled to be useful for the purpose of computation.

At present, no single reason behind the speedup was found from the
standpoint of entanglement and discord. The speedup appears to always depend
on the exact nature of the problem while the reason for it varies from
problem to problem $\left[ 22\right] $. The relation between number of
oracle queries and quantum retocausality highlighted in the present work
appears to hold exactly for any oracle problem.

We should also cite tree size complexity $\left[ 24\right] $ and
contextually based $\left[ 25\right] $ arguments. In the former, a measure
of the complexity of the multiqubit state is shown to be related to the
speedup of a variety of quantum algorithms. The latter addresses the
relation between speedup and the contextual character of quantum mechanics.
It identifies a form of fault tolerant quantum computation (by \textit{magic
states}) specially resilient to noise. Also the present retrocausal
explanation of the speedup could be considered a contextually based
argument. The reduction of the initial and final measurements of a quantum
process to partial non-redundant measurements is of course contextual in
character.

\section{The seminal Deutsch algorithm}

Let us review the usual representation of Deutsch algorithm, limited to the
process of solving the problem. We need two quantum registers: $A$, of basis
vectors $\left\vert 0\right\rangle _{A}$ and $\left\vert 1\right\rangle _{A}$%
, and $V$, of basis vectors $\left\vert 0\right\rangle _{V}$ and $\left\vert
1\right\rangle _{V}$. We use ket vectors instead of density operators as in
the original Deutsch algorithm.

Bob chooses one of the four functions in array (\ref{deutsch}), say $%
f_{01}\left( \mathbf{a}\right) $, and gives Alice the black box that
computes it. Alice knows array (\ref{deutsch}) but does not know which is
the function chosen by Bob. She is to find whether it is constant or
balanced through function evaluations. She prepares register $A$ with the
value of $\mathbf{a}$ for which she wants to perform function evaluation.
The black box computes the value of $f_{01}\left( \mathbf{a}\right) $ and
adds it module two to the former content of register $V$. Being logically
reversible, module two addition can be implemented unitarily. In the
introduction we omitted register $V$ because transformations are unitary
also without it, but they are more difficult to explain.

For reasons that will soon become clear, the input state of the quantum
algorithm is: 
\begin{equation*}
\left\vert \psi \right\rangle =\frac{1}{\sqrt{2}}\left\vert 0\right\rangle
_{A}\left( \left\vert 0\right\rangle _{V}-\left\vert 1\right\rangle
_{V}\right) .
\end{equation*}

Alice applies to register $A$ the Hadamard transform $H_{A}$, which
transforms $\left\vert 0\right\rangle _{A}$ into $\frac{1}{\sqrt{2}}\left(
\left\vert 0\right\rangle _{A}+\left\vert 1\right\rangle _{A}\right) $ and $%
\left\vert 1\right\rangle _{A}$ into $\frac{1}{\sqrt{2}}\left( \left\vert
0\right\rangle _{A}-\left\vert 1\right\rangle _{A}\right) $, producing the
state: 
\begin{equation}
H_{A}\left\vert \psi \right\rangle =\frac{1}{2}\left( \left\vert
0\right\rangle _{A}+\left\vert 1\right\rangle _{A}\right) \left( \left\vert
0\right\rangle _{V}-\left\vert 1\right\rangle _{V}\right) ,  \label{ex}
\end{equation}%
then asks the black box to compute the value of the function. Let $U_{f}$ be
the corresponding unitary transformation (defined in the Hilbert space of
all registers). We have:%
\begin{equation*}
U_{f}H_{A}\left\vert \psi \right\rangle =\frac{1}{2}\left( \left\vert
0\right\rangle _{A}-\left\vert 1\right\rangle _{A}\right) \left( \left\vert
0\right\rangle _{V}-\left\vert 1\right\rangle _{V}\right) .
\end{equation*}

Function evaluation is performed in quantum parallelism for each term of the
input state superposition. It leaves the term $\left\vert 0\right\rangle
_{A}\left( \left\vert 0\right\rangle _{V}-\left\vert 1\right\rangle
_{V}\right) $, appearing in the input state (\ref{ex}), unaltered. In fact,
here the argument of the function, the content of register $A$, is $0$. The
computation of $f_{01}\left( 0\right) $ yields $0$ that module two added to
the former content of register $V$ leaves everything unaltered. Function
evaluation instead changes the term $\left\vert 1\right\rangle _{A}\left(
\left\vert 0\right\rangle _{V}-\left\vert 1\right\rangle _{V}\right) $ into $%
\left\vert 1\right\rangle _{A}\left( \left\vert 1\right\rangle
_{V}-\left\vert 0\right\rangle _{V}\right) =-\left\vert 1\right\rangle
_{A}\left( \left\vert 0\right\rangle _{V}-\left\vert 1\right\rangle
_{V}\right) $. In fact now we have to module two add $f_{01}\left( 1\right)
=1$ and this changes $\left\vert 0\right\rangle _{V}$ into $\left\vert
1\right\rangle _{V}$ and $\left\vert 1\right\rangle _{V}$ into $\left\vert
0\right\rangle _{V}$.

Then Alice applies a second time the Hadamard transform to register $A$,
obtaining the output state:%
\begin{equation}
H_{A}U_{f}H_{A}\left\vert \psi \right\rangle =\frac{1}{\sqrt{2}}\left\vert
1\right\rangle _{A}\left( \left\vert 0\right\rangle _{V}-\left\vert
1\right\rangle _{V}\right) .  \label{outu}
\end{equation}

Eventually she measures the \textit{content} of register $A$, namely the
observable $\hat{A}$ of eigenstates $\left\vert 0\right\rangle _{A}$ and $%
\left\vert 1\right\rangle _{A}$ and eigenvalues respectively $0$ and $1$.
She reads the eigenvalue $1$, which tells her that the function is balanced
(the final content of register $A$ is $0$ when the function is constant and $%
1$ when it is balanced).

Thus the problem of checking whether the function given by Bob is constant
or balanced is always solved with just one function evaluation quantumly,
against two classically.

The mathematics of this speedup, namely that of the quantum algorithm, is
obvious in the sense that we have it under the eyes. However, the
mathematics of different quantum algorithms are different from one another
as there is no known universal scheme. The \textit{mechanism} of the
speedups, provided there is one, is not known.

\subsection{Time-symmetric and relativized representations}

To start with, we extend the representation of Deutsch algorithm to the
process of choosing the black box. To this end, we should add an imaginary
quantum register $B$ of basis vectors $\left\vert 00\right\rangle _{B}$, $%
\left\vert 01\right\rangle _{B}$, $\left\vert 10\right\rangle _{B}$, and $%
\left\vert 11\right\rangle _{B}$. This register contains the problem
setting, namely the suffix $\mathbf{b}$\ of the function chosen by Bob. The
previous black box, which computed $f_{\mathbf{b}}\left( \mathbf{a}\right) $%
\ for a well determined value of $\mathbf{b}$ and any value of $\mathbf{a}$%
,\ is replaced by a universal one that computes $f_{\mathbf{b}}\left( 
\mathbf{a}\right) $ for any values of $\mathbf{b}$ and $\mathbf{a}$.
Register $A$ and $V$ have the same role as before.

For reasons that will soon become clear, we assume that register $B$ is
initially in the maximally mixed state: 
\begin{equation*}
\rho _{B}=\frac{1}{4}\left( \left\vert 00\right\rangle \left\langle
00\right\vert _{B}+\left\vert 01\right\rangle \left\langle 01\right\vert
_{B}+\left\vert 10\right\rangle \left\langle 10\right\vert _{B}+\left\vert
11\right\rangle \left\langle 11\right\vert _{B}\right) .
\end{equation*}

As we will need a detailed representation of quantum states and operators,
for reasons of encumbrance we represent all states as ket vectors, not
matrices. To this end, we move to the random phase representation $\left[ 26%
\right] $ of the maximally mixed state of register $B$:

\begin{equation}
\left\vert \psi \right\rangle _{B}=\frac{1}{2}\left( \func{e}^{i\varphi
_{0}}\left\vert 00\right\rangle _{B}+\func{e}^{i\varphi _{1}}\left\vert
01\right\rangle _{B}+\func{e}^{i\varphi _{2}}\left\vert 10\right\rangle _{B}+%
\func{e}^{i\varphi _{3}}\left\vert 11\right\rangle _{B}\right) ,  \label{ini}
\end{equation}%
where the $\varphi _{i}$ are independent random phases with uniform
distribution in $\left[ 0,2\pi \right] $. We will be dealing with a trivial
application of the random phase representation: we can always think that the
quantum state evolves as a pure state with the $\varphi _{i}$ fixed phases.
Only when we have to compute its von Neumann entropy, we should remember
that the $\varphi _{i}$ are random variables. The von Neumann entropy of
state (\ref{ini}), as that of $\rho _{B}$, is two bit.

By the way, $\rho _{B}$ is the average over all $\varphi _{i}$ of the
product of the ket by the bra: $\rho _{B}=\left\langle \left\vert \psi
\right\rangle _{B}\left\langle \psi \right\vert _{B}\right\rangle _{\forall
\varphi _{i}}$; reading state (\ref{ini}) is also simple: it is a mixture of
pure states with the phases $\varphi _{0}$, $\varphi _{1}$, $\varphi _{2}$, $%
\varphi _{3}$ all different, in fact a dephased quantum superposition.$\ $

The overall initial state of the three registers, at time $t_{0}$, is thus:%
\begin{equation}
\left\vert \psi \right\rangle =\frac{1}{2\sqrt{2}}\left( \func{e}^{i\varphi
_{0}}\left\vert 00\right\rangle _{B}+\func{e}^{i\varphi _{1}}\left\vert
01\right\rangle _{B}+\func{e}^{i\varphi _{2}}\left\vert 10\right\rangle _{B}+%
\func{e}^{i\varphi _{3}}\left\vert 11\right\rangle _{B}\right) \left\vert
0\right\rangle _{A}\left( \left\vert 0\right\rangle _{V}-\left\vert
1\right\rangle _{V}\right) .  \label{in}
\end{equation}

In order to prepare register $B$ in the desired problem setting, at time\ $%
t_{0}$ Bob measures its content, namely the observable $\hat{B}$ of
eigenstates the basis vectors $\left\vert 00\right\rangle _{B},\left\vert
01\right\rangle _{B},...$ and eigenvalues respectively $00,01,...$. Note
that $\hat{B}$ commutes with $\hat{A}$.\ The measurement outcome is
completely random. Say it comes out the eigenvalue $\mathbf{b}=10$. The
state immediately after measurement is:

\begin{equation}
P_{B}\left\vert \psi \right\rangle =\frac{1}{\sqrt{2}}\left\vert
10\right\rangle _{B}\left\vert 00\right\rangle _{A}\left( \left\vert
0\right\rangle _{V}-\left\vert 1\right\rangle _{V}\right) ,  \label{pro}
\end{equation}%
where $P_{B}$ is the projection of the quantum state induced by Bob's
measurement. Then Bob applies to register $B$ a unitary transformation $%
U_{B} $ that changes the random measurement outcome into the desired problem
setting, say $\mathbf{b}=01$. At time $t_{1}$ we will have: 
\begin{equation}
U_{B}P_{B}\left\vert \psi \right\rangle =\frac{1}{\sqrt{2}}\left\vert
01\right\rangle _{B}\left\vert 0\right\rangle _{A}\left( \left\vert
0\right\rangle _{V}-\left\vert 1\right\rangle _{V}\right) .  \label{inbob}
\end{equation}

State (\ref{inbob}) is the input state of the quantum algorithm in the
representation extended to the process of setting the problem. There are of
course many $U_{B}$ that change $\left\vert 10\right\rangle _{B}$ into $%
\left\vert 01\right\rangle _{B}$. For simplicity of exposition, we choose
the one that bit by bit changes zeros into ones and ones into zeros:%
\begin{equation*}
U_{B}\equiv \left\vert 11\right\rangle \left\langle 00\right\vert
_{B}+\left\vert 10\right\rangle \left\langle 01\right\vert _{B}+\left\vert
01\right\rangle \left\langle 10\right\vert _{B}+\left\vert 00\right\rangle
\left\langle 11\right\vert _{B}.
\end{equation*}

The output state of the extended representation of the quantum algorithm is:%
\begin{equation}
H_{A}U_{f}H_{A}U_{B}P_{B}\left\vert \psi \right\rangle =\frac{1}{\sqrt{2}}%
\left\vert 01\right\rangle _{B}\left\vert 1\right\rangle _{A}\left(
\left\vert 0\right\rangle _{V}-\left\vert 1\right\rangle _{V}\right) .
\label{outbob}
\end{equation}%
Of course, input and output states are the same as in the usual
representation of the quantum algorithm up to the presence of the ket $%
\left\vert 01\right\rangle _{B}$.

We note that this extension immediately calls for another one, this time
concerning the actors (observers) on the stage. We have to resort to the
relational quantum mechanics of Rovelli $\left[ 10\right] $, where quantum
states are observer dependent. State (\ref{inbob}) is with respect to Bob,
the problem setter, and any other observer who does not act on the problem
solving process. It cannot be with respect to Alice, the problem solver. The
sharp state $\left\vert 01\right\rangle _{B}$ would tell her, before she
starts her search for the solution, that the function chosen by Bob is $%
f_{01}\left( \mathbf{a}\right) $. She would know that it is balanced without
performing any function evaluation. Of course the suffix of the function
should be hidden to Alice -- to her it is inside the black box.

To physically represent this fact, it suffices to retard the projection $%
P_{B}$\ until the end of the unitary part of Alice's action, at time $t_{2}$.

To her, the state of register $B$ in the input state of the quantum
algorithm is still maximally mixed. In fact $U_{B}$ leaves state (\ref{in})
unaltered up to an irrelevant permutation of the independent random phases.
Thus, disregarding the permutation, state (\ref{in}) is the input state to
Alice.

We started with register $B$ in a maximally mixed state to represent the
fact that, to Alice, the problem setting is physically hidden.

Summing up, states (\ref{in}) through (\ref{outbob}) are the representation
of the quantum algorithm with respect to Bob. In the representation with
respect to Alice, the input state, which coincides with the initial state,
is: 
\begin{equation}
U_{B}\left\vert \psi \right\rangle =\left\vert \psi \right\rangle =\frac{1}{2%
\sqrt{2}}\left( \func{e}^{i\varphi _{0}}\left\vert 00\right\rangle _{B}+%
\func{e}^{i\varphi _{1}}\left\vert 01\right\rangle _{B}+\func{e}^{i\varphi
_{2}}\left\vert 10\right\rangle _{B}+\func{e}^{i\varphi _{3}}\left\vert
11\right\rangle _{B}\right) \left\vert 0\right\rangle _{A}(\left\vert
0\right\rangle _{V}-\left\vert 1\right\rangle _{V}).  \label{ina}
\end{equation}%
The two bit entropy of the state of register $B$ represents Alice's complete
ignorance of the problem setting. The output state is:

\begin{eqnarray}
H_{A}U_{f}H_{A}U_{B}\left\vert \psi \right\rangle &=&\frac{1}{2\sqrt{2}}%
\left[ \left( \func{e}^{i\varphi _{0}}\left\vert 00\right\rangle _{B}+\func{e%
}^{i\varphi _{3}}\left\vert 11\right\rangle _{B}\right) \left\vert
0\right\rangle _{A}+\left( \func{e}^{i\varphi _{1}}\left\vert
01\right\rangle _{B}-\func{e}^{i\varphi _{2}}\left\vert 10\right\rangle
_{B}\right) \left\vert 1\right\rangle _{A}\right]  \notag \\
&&(\left\vert 0\right\rangle _{V}-\left\vert 1\right\rangle _{V}),
\label{outa}
\end{eqnarray}%
We can see that, for each possible problem setting (value of $\mathbf{b}$
contained in register $B$), Alice has built the corresponding solution of
the problem $s\left( \mathbf{b}\right) $\ in register $A$.

Eventually, at time $t_{2}$, she acquires the solution by reading the
content of register $A$, namely by measuring $\hat{A}$. We should keep in
mind that the output state (\ref{outa}) is with respect to Alice. The same
state with respect to Bob and any other observer is $\frac{1}{\sqrt{2}}%
\left\vert 01\right\rangle _{B}\left\vert 1\right\rangle _{A}\left(
\left\vert 0\right\rangle _{V}-\left\vert 1\right\rangle _{V}\right) $. The
measurement outcome is unpredictable to Alice, it is already $1$ to any
other observer. Thus Alice's measurement must select the eigenvalue $1$ with
probability one, projecting state (\ref{outa}) on%
\begin{equation}
P_{A}H_{A}U_{f}H_{A}U_{B}\left\vert \psi \right\rangle =\frac{1}{\sqrt{2}}%
\left( \func{e}^{i\varphi _{1}}\left\vert 01\right\rangle _{B}-\func{e}%
^{i\varphi _{2}}\left\vert 10\right\rangle _{B}\right) \left\vert
1\right\rangle _{A}\left( \left\vert 0\right\rangle _{V}-\left\vert
1\right\rangle _{V}\right) ,  \label{alice}
\end{equation}%
where $P_{A}$\ is the projection induced by the final Alice's measurement.
State (\ref{alice}) is further projected on: 
\begin{equation}
\frac{1}{\sqrt{2}}\left\vert 01\right\rangle _{B}\left\vert 1\right\rangle
_{A}\left( \left\vert 0\right\rangle _{V}-\left\vert 1\right\rangle
_{V}\right)  \label{outb}
\end{equation}%
by the retarded projection induced by the initial Bob's measurement. We note
that inverting the order of the two projections leaves the end result
unaltered. As a matter of fact, since the projection due to Bob's
measurement cannot be retarded beyond the unitary part of Alice's action, we
should see the two projections as simultaneous. In this way Alice, by
measuring $\hat{A}$, also acquires the content of register $B$. In fact
state (\ref{outb}), with register $B$ in the sharp state $\left\vert
01\right\rangle _{B}$, tells Alice that the problem setting chosen by Bob is 
$\mathbf{b}=01$. Note that this state is common to both representations, to
Alice and to Bob.

In view of what will follow, we note that Alice's measurement of $\hat{A}$\
in the output state relativized to her is equivalent to the measurement of $%
\hat{B}$. In fact either measurement projects state (\ref{outa}) on (\ref%
{outb}), where the sharp states of registers $B$ and $A$ tell Alice both the
setting and the solution of the problem.

\subsection{Quantum feedback}

We consider the random phase representation of the reduced density operator
of register $A$ in the output state (\ref{outa}):%
\begin{equation}
\left\vert \psi \right\rangle _{A}=\frac{1}{\sqrt{2}}\left( \func{e}^{i\Phi
_{0}}\left\vert 0\right\rangle _{A}+\func{e}^{i\Phi _{1}}\left\vert
1\right\rangle _{A}\right) ,  \label{reduced}
\end{equation}%
\ where $\Phi _{0}$ and $\Phi _{1}$ are independent random phases with
uniform distribution in $\left[ 0,2\pi \right] $. The usual representation
is $\rho _{A}=\frac{1}{2}\left( \left\vert 0\right\rangle _{A}\left\langle
0\right\vert _{A}+\left\vert 1\right\rangle _{A}\left\langle 1\right\vert
_{A}\right) $.

$\mathcal{E}_{A}$, the entropy of $\left\vert \psi \right\rangle _{A}$, is $%
1 $ bit.\ The zeroing of $\mathcal{E}_{A}$ can be due to either the
projection of the quantum state associated with the measurement of $\hat{B}$
in the initial state (\ref{in}), retarded at the end of the unitary part of
Alice's action, or that associated with the measurement of $\hat{A}$ in the
output state (\ref{outa}) (we have seen that the two projections should be
considered simultaneous). The present work is an exploration of the
assumption that the zeroing of $\mathcal{E}_{A}$ shares between the two
measurements.

To this end, we assume that the two complete measurements reduce to partial
measurements such that: \textbf{1} together, they select whatever was
selected by the complete measurements and \textbf{2} each performed alone,
contribute in a complementary and non-redundant way to the zeroing of $%
\mathcal{E}_{A}$. By this we mean that no information provided by either
partial measurement is provided by the other.

We call \textbf{1} and \textbf{2} \textit{Occam conditions}. They can be
seen as an application of Occam razor. In Newton's formulation, it states: 
\textit{We are to admit no more causes of natural things than such that are
both true and sufficient to explain their appearances}.\ $\left[ 27\right] $%
. Here, the razor should exclude any redundancy between initial and final
measurement.

The assumption that the two partial measurements contribute equally to the
zeroing of $\mathcal{E}_{A}$, namely that $R=\frac{1}{2}$, explains the
speedup of the present quantum algorithm.

We should reduce the initial Bob's measurement and the final Alice's
measurement to two partial measurements submitted to \textbf{1} and \textbf{2%
} and the condition of equally contributing to the zeroing of $\mathcal{E}%
_{A}$.

We have seen that the measurement of $\hat{A}$ in the relativized output
state (\ref{outa}) is equivalent to that of $\hat{B}$. Thus we can move to
the problem of reducing two measurements of $\hat{B}$, one performed by Bob
in the initial state (\ref{in}) and the other by Alice in the output state (%
\ref{outa}), to two partial measurements, say of $B_{i}$ and $\hat{B}_{j}$,
satisfying the above said conditions. In the most general terms, $B_{i}$ and 
$\hat{B}_{j}$ are Boolean functions of $\hat{B}$, such as: $\hat{B}_{0}$,
the content of the left cell of register $B$, $\hat{B}_{1}$, the content of
the right cell, $\limfunc{XOR}\left( \hat{B}_{0},\hat{B}_{1}\right) $, the
exclusive or between the two former contents, etc.

We provide an example of reduction of the complete measurements to such
partial measurements. We keep the assumption that the initial measurement of 
$\hat{B}$ randomly selects the eigenvalue $\mathbf{b}=10$\ and that Bob, by $%
U_{B}$, changes it into $\mathbf{b}=01$. Let $\mathbf{b}\equiv b_{0}b_{1}$;
we assume that the eigenvalue $b_{0}=1$\ is selected at time $t_{0}$ by the
measurement of $\hat{B}_{0}$ in the initial state\ and that the eigenvalue $%
b_{1}=1$ is selected at time $t_{2}$ by the measurement of $\hat{B}_{1}$ in
the output state.

To reconstruct the selections performed by the complete measurements, we
should propagate forward in time, by $H_{A}U_{f}H_{A}U_{B}$, the projection
induced by the former measurement and backward in time, by its inverse, the
projection induced by the latter measurement. The two propagations can be
performed in any order, the reconstruction is the same.

Let us perform the backward propagation first. The measurement of $B_{1}$\
in the output state (\ref{outa}), which assumedly selects $b_{1}=1$,
projects this state on: 
\begin{equation}
\left\vert \chi \right\rangle =\frac{1}{2}\left( \func{e}^{i\varphi
_{1}}\left\vert 01\right\rangle _{B}\left\vert 1\right\rangle _{A}+\func{e}%
^{i\varphi _{3}}\left\vert 11\right\rangle _{B}\left\vert 0\right\rangle
_{A}\right) \left( \left\vert 0\right\rangle _{V}-\left\vert 1\right\rangle
_{V}\right) .  \label{co}
\end{equation}

We advance at time $t_{0}$ the two ends of this projection. The result is
the projection of the initial state (\ref{in})\ on:%
\begin{equation}
U_{B}^{\dag }H_{A}^{\dag }U_{f}^{\dag }H_{A}^{\dag }\left\vert \chi
\right\rangle =\frac{1}{2}\left( \func{e}^{i\varphi _{3}}\left\vert
00\right\rangle _{B}+\func{e}^{i\varphi _{1}}\left\vert 10\right\rangle
_{B}\right) \left\vert 0\right\rangle _{A}(\left\vert 0\right\rangle
_{V}-\left\vert 1\right\rangle _{V}).  \label{mo}
\end{equation}

The permutation of the independent random phases is irrelevant. At this
point the measurement of $\hat{B}_{0}$ in state (\ref{mo}), which assumedly
selects $b_{0}=1$, projects it on:%
\begin{equation}
\left\vert \xi \right\rangle =\frac{1}{\sqrt{2}}\left\vert 10\right\rangle
_{B}\left\vert 0\right\rangle _{A}\left( \left\vert 0\right\rangle
_{V}-1_{V}\right) .  \label{dino}
\end{equation}

Of course state (\ref{dino}), under $H_{A}U_{f}H_{A}U_{B}$, evolves into
state (\ref{outb}), namely $\frac{1}{\sqrt{2}}\left\vert 01\right\rangle
_{B}\left\vert 1\right\rangle _{A}\left( \left\vert 0\right\rangle
_{V}-\left\vert 1\right\rangle _{V}\right) $, the final state common to both
representations (to Bob and to Alice). We have reconstructed the selections
performed by the complete measurements. Furthermore, the reduction of $%
\mathcal{E}_{A}$ induced by either partial measurement, performed alone, is
half bit and no information acquired by either partial measurement is
acquired by the other. Conditions \textbf{1} and \textbf{2} are satisfied.

One can see that, eventually, everything boils down to ascribing the
selection of one of the two bits (the right one in present assumptions) of
the random outcome of the initial measurement to the final measurement. We
are not sending a message backward in time. Each of the bits that specify
the outcome of the initial measurement is independently and randomly
selected. We are just ascribing half of these random selections to the final
rather than the initial measurement.

We note that sharing between Bob's and Alice's measurements the zeroing of $%
\mathcal{E}_{A}$ does not affect Bob's freedom of choosing the function
computed by the black box. We should keep in mind that the probability that
Alice's measurement of $\hat{B}$\ in state (\ref{outa}) selects $\mathbf{b}%
=01$, or that the measurement of $\hat{B}_{1}$ selects $b_{1}=1$ (the right
digit of $01$), is one. This means that the measurement of $\hat{B}_{1}$\
just reads the right digit of the problem setting $\mathbf{b}=01$\ freely
chosen (determined) by Bob, without possibly altering it, or affecting Bob's
freedom of choosing it. This goes along with the fact that the backward
propagation of the projection due to the measurement of $\hat{B}_{1}$ in the
output state does not determine any part of Bob's choice, but the right
digit of the random outcome of Bob's measurement $\mathbf{b}=10$, which is
before that choice.

The kind of\textit{\ }retrocausation discussed above is sometimes invoked to
explain EPR non-locality, but mostly as a curiosity because it is believed
to be of no consequence. It has no consequences also in the representation
of the quantum algorithm with respect to Bob and any external observer. To
them, it leaves the input state of the algorithm -- state (\ref{inbob}) --\
unaltered. It just tells that, say, the left digit of the random outcome of
Bob's measurement $\mathbf{b}=10$ has been randomly selected by Bob's
measurement and the right digit has been randomly selected back in time by
the future Alice's measurement -- in fact an inconsequential thing.

Things change dramatically in the representation with respect to Alice --
the problem solver.

We have seen that the projection induced by Alice's measurement of $\hat{B}%
_{1}$ in the output state (\ref{outa}) must propagate backward in time
through the inverse of $H_{A}U_{f}H_{A}U_{B}$ until $t_{0}$, where it
selects the right digit of the random outcome of Bob's measurement $10$.\
Let us see the value of this backward propagation at time $t_{1}$,
immediately after the application of $U_{B}$ and before that of $%
H_{A}U_{f}H_{A}$. This time we should advance the two ends of the projection
of state (\ref{outa}) on state (\ref{co}) by the inverse of $H_{A}U_{f}H_{A}$%
. The result is the projection of state (\ref{ina}),\ the input state of the
quantum algorithm with respect to Alice, on:%
\begin{equation}
H_{A}^{\dag }U_{f}^{\dag }H_{A}^{\dag }\left\vert \chi \right\rangle =\frac{1%
}{2}\left( \func{e}^{i\varphi _{1}}\left\vert 01\right\rangle _{B}+\func{e}%
^{i\varphi _{3}}\left\vert 11\right\rangle _{B}\right) \left\vert
0\right\rangle _{A}\left( \left\vert 0\right\rangle _{V}-\left\vert
1\right\rangle _{V}\right) .  \label{adv}
\end{equation}

This is an outstanding consequence. State (\ref{adv}), the input state to
Alice under the assumption that the selection of the solution equally shares
between Bob's and Alice's measurements,\ tells her, before she performs any
function evaluation, that the suffix of the function chosen by Bob is either 
$\mathbf{b}=01$ or $\mathbf{b}=11$, namely that $\mathbf{b}\in \left\{
01,11\right\} $. We can say that Alice \textit{knows in advance} that $%
\mathbf{b}\in \left\{ 01,11\right\} $, since this knowledge comes from the
projection of the quantum state induced by her future measurement.

We provide the following interpretation of this advanced knowledge. We are
at a fundamental level where knowing is doing $\left[ 28\right] $. Alice is
the problem solver, her knowing in advance that $\mathbf{b}\in \left\{
01,11\right\} $ would simply mean that the quantum algorithm requires the
number of function evaluations logically required to identify the solution
starting from that knowledge. We mean by classical logic. This of course
establishes a correspondence between quantum computation and classical
logic. It is the main assumption of the present work.

In the present case, the number of function evaluations required to
discriminate between $f_{01}\left( \mathbf{a}\right) $ and $f_{11}\left( 
\mathbf{a}\right) $ is just one. In fact the value of the function for the
argument $\mathbf{a}=0$ does the job -- see the tables of the two functions
in question in array (\ref{deutsch}). Since it is $0$, the function must be $%
f_{01}\left( \mathbf{a}\right) $ This implies that it is balanced.

With problem setting $\mathbf{b}=01$, there are three instances of Alice's
advanced knowledge: $\mathbf{b}\in \left\{ 01,00\right\} _{B}$, $\mathbf{b}%
\in \left\{ 01,11\right\} _{B}$, and $\mathbf{b}\in \left\{ 01,10\right\}
_{B}$. As one can see, $\mathbf{b}=01$ goes with each one of the other three
possible values of $\mathbf{b}$.  Similarly for problem setting $\mathbf{b}%
=00$, etc.  It turns out that the quantum algorithm can be seen as a sum
over classical histories in each of which Alice knows in advance that the
value of $\mathbf{b}$\ chosen by Bob is either one of a particular pair of
values and performs the function evaluation for the value of the argument
that tells which one.

Let us see this in more detail. We see the quantum algorithm under the
perspective of Feynman's sum over classical histories $\left[ 29\right] $. A
classical history is a classical trajectory of the quantum registers, namely
a causal sequence of sharp register states. For example: 
\begin{equation}
\func{e}^{i\varphi _{1}}\left\vert 01\right\rangle _{B}\left\vert
0\right\rangle _{A}\left\vert 0\right\rangle _{V}\overset{H_{A}}{\rightarrow 
}\func{e}^{i\varphi _{1}}\left\vert 01\right\rangle _{B}\left\vert
0\right\rangle _{A}\left\vert 0\right\rangle _{V}\overset{U_{f}}{\rightarrow 
}\func{e}^{i\varphi _{1}}\left\vert 01\right\rangle _{B}\left\vert
0\right\rangle _{A}\left\vert 0\right\rangle _{V}\overset{H_{A}}{\rightarrow 
}\func{e}^{i\varphi _{1}}\left\vert 01\right\rangle _{B}\left\vert
1\right\rangle _{A}\left\vert 0\right\rangle _{V}.  \label{one}
\end{equation}

The left-most state is one of the elements of the input\ state superposition
(\ref{ina}). The state after each arrow is one of the elements of the
superposition generated by the unitary transformation of the state before
the arrow; the transformation in question is specified above the arrow.

In history (\ref{one}), the problem setting is $\mathbf{b}=01$. Alice
performs function evaluation for $\mathbf{a}=0$ (second and third state).
This behavior is justifiable by two instances of Alice's advanced knowledge.
One is $\mathbf{b}\in \left\{ 01,11\right\} _{B}$, the other $\mathbf{b}\in
\left\{ 01,10\right\} _{B}$. The value of the function for $\mathbf{a}=0$ in
either case tells that the function in the black box is $f_{01}\left( 
\mathbf{a}\right) $ and thus that it is balanced.

Summing up, with $R=\frac{1}{2}$, the quantum algorithm can be seen as a sum
over classical histories in each of which Alice knows in advance one of the
possible halves of the problem setting and performs the function evaluations
logically required to identify the missing half and thus the solution.

\section{Generalization}

We show that, given an oracle problem, we can know the number of function
evaluations required to solve it with quantum retrocausality $R=\frac{1}{2}$%
. A generic oracle problem can be formulated as follows. We have a set of
functions $f_{\mathbf{b}}:\left\{ 0,1\right\} ^{n}\rightarrow \left\{
0,1\right\} ^{m}$ with $m\leq n$. The suffix $\mathbf{b}$ ranges over the
set of all the problem settings $\sigma _{B}$. Bob chooses one of these
functions (a value of $\mathbf{b}$) and gives Alice the black box (oracle)
that computes it. Alice knows the set of functions but does not know which
is the function chosen by Bob. She is to find a certain feature of the
function (eg whether it is constant or balanced in the algorithm of Deutsch,
or its period in that of Shor) by performing function evaluations (oracle
queries). We call the feature in question, which is the solution of the
problem and a function of $\mathbf{b}$, $s\left( \mathbf{b}\right) $.

\subsection{Time-symmetric representation to Alice}

Provided that a register $B$ contains the problem setting $\mathbf{b}$ and a
register $A$ will eventually contain the solution $s\left( \mathbf{b}\right) 
$, the most general form of the input and output states of the unitary part (%
$U$) of Alice's problem-solving action, in the representation of the quantum
algorithm to her, is:%
\begin{align}
\left\vert \func{in}\right\rangle _{BAW}& =\frac{1}{\sqrt{c}}\left(
\dsum\limits_{\mathbf{b~\in ~}\sigma _{B}}\func{e}^{i\varphi _{\mathbf{b}%
}}\left\vert \mathbf{b}\right\rangle _{B}\right) \left\vert
00...\right\rangle _{A}\left\vert \psi \right\rangle _{W},  \label{input} \\
\left\vert \func{out}\right\rangle _{BAW}& =U\left\vert \func{in}%
\right\rangle _{BAW}=\frac{1}{\sqrt{c}}\dsum\limits_{\mathbf{b~\in ~}\sigma
_{B}}\func{e}^{i\varphi _{\mathbf{b}}}\left\vert \mathbf{b}\right\rangle
_{B}\left\vert s\left( \mathbf{b}\right) \right\rangle _{A}\left\vert
\varphi \left( \mathbf{b}\right) \right\rangle _{W},  \label{output}
\end{align}%
where $c$ is the cardinality of $\sigma _{B}$, $\left\vert \psi
\right\rangle _{W}$ and $\left\vert \varphi \left( \mathbf{b}\right)
\right\rangle _{W}\ $are normalized states$\ $of a register $W$, which
stands for any other register or set of registers.

$U$ should not change the problem setting. It suffices that register $B$ is
the control register of all function evaluations, what means that the
content of register $B$ affects the output of function evaluation while
remaining unaltered through it, and the unitary transformations before and
after each function evaluation do not apply to $B$. Correspondingly, $U$\
sends the input into the output independently term by term and keeping the
value of $\mathbf{b}$\ unaltered:

\begin{equation}
\forall \mathbf{b:~~}U\left\vert \mathbf{b}\right\rangle _{B}\left\vert
00...\right\rangle _{A}\left\vert \psi \right\rangle _{W}=\left\vert \mathbf{%
b}\right\rangle _{B}\left\vert s\left( \mathbf{b}\right) \right\rangle
_{A}\left\vert \varphi \left( \mathbf{b}\right) \right\rangle _{W}.
\label{indi}
\end{equation}

Given the oracle problem, namely all the pairs $\mathbf{b}$ and $s\left( 
\mathbf{b}\right) $, and provided that one is free to add suitable \textit{%
garbage qubits} to register $W$, it should not be difficult to put the input
and output states in a form compatible with the existence of such a $U$
between them. In the following, we assume that states (\ref{input}) and (\ref%
{output}) are of this form. We will see that we do not need to know the form
of $U$ to the end of ascertaining the number of function evaluations
required to solve the oracle problem with quantum retrocausality $R=\frac{1}{%
2}$; it suffices to know all the pairs $\mathbf{b}$ and $s\left( \mathbf{b}%
\right) $.

Note that, for equation (\ref{indi}), the projection of the quantum state
induced by any measurement on the content of register $B$ in the output
state, advanced by $U^{\dag }$, becomes the projection induced by performing
the same measurement in the input state. Conversely, the projection induced
by any measurement on the content of $B$ in the input state, retarded by $U$%
, becomes the projection induced by performing the same measurement in the
output state. This goes along with the fact that the reduced density
operator of register $B$ remains the same throughout $U$. Its random phase
representation is%
\begin{equation}
\left\vert \psi \right\rangle _{B}=\frac{1}{\sqrt{c}}\dsum\limits_{\mathbf{%
b~\in ~}\sigma _{B}}\func{e}^{i\varphi _{\mathbf{b}}}\left\vert \mathbf{b}%
\right\rangle _{B},  \label{red}
\end{equation}%
the usual representation being $\rho _{B}=\frac{1}{c}\dsum\limits_{\mathbf{%
b~\in ~}\sigma _{B}}\left\vert \mathbf{b}\right\rangle \left\langle \mathbf{b%
}\right\vert _{B}$.

\subsection{Quantum feedback}

Given the equations (\ref{input}) and (\ref{output}), we show how to share
the selection of the solution between initial and final measurements and
derive the corresponding Alice's advanced knowledge.

It is simpler to assume that $U_{B}$ is the identity. In this way we can
think that the initial Bob's measurement is performed in state (\ref{input}%
). Of course its selection of a value of $\mathbf{b}$ also determines that
of $s\left( \mathbf{b}\right) $.

We reformulate Occam conditions \textbf{1} and \textbf{2} for the particular
case $R=\frac{1}{2}$. We should reduce in all the possible ways the two
measurements of $\hat{B}$, one on the part of Bob in the input state and the
other on the part of Alice in the output state (see Section 4.2)\ to two
partial measurements -- of $\hat{B}_{i}$ and $\hat{B}_{j}$ -- such that:

\begin{enumerate}
\item[\textbf{I}] together, they select whatever is selected by the complete
measurements and

\item[\textbf{II}] each performed alone, they contribute in an equal and
non-redundant way to the selection of the solution.
\end{enumerate}

Let $\mathcal{E}_{A}$\ be the von Neumann entropy of the solution, namely of
the trace over registers $B$ and $W$ of state $\left\vert \func{out}%
\right\rangle _{BAW}$. Point \textbf{II}\ implies the following two
conditions:%
\begin{equation}
\Delta \mathcal{E}_{A}\left( \hat{B}_{i}\right) =\Delta \mathcal{E}%
_{A}\left( \hat{B}_{j}\right) ,  \label{equg}
\end{equation}%
where $\Delta \mathcal{E}_{A}\left( \hat{B}_{i}\right) $ is the reduction of 
$\mathcal{E}_{A}$ due to the measurement of $\hat{B}_{i}$, $\Delta \mathcal{E%
}_{A}\left( \hat{B}_{j}\right) $ that due to the measurement of $\hat{B}_{j}$%
, and:%
\begin{equation}
\fbox{$no~partial~measurement~outcome\
provides~enough~information~to~select~the~solution\text{.}$}\text{~}
\label{no}
\end{equation}%
In fact the cases are two: if both outcomes provided enough information,
then there would be redundant information, what is forbidden by the
no-redundancy condition. If only one did, then the two partial measurements
would not contribute equally to the selection of the solution, what is
forbidden by the equality condition. Condition (\ref{no}) is redundant when $%
\mathbf{b}$ is an unstructured bit string as in Deutsch algorithm, it is not
when $\mathbf{b}$ is structured.

Alice's measurement of $\hat{B}_{j}$ (as any measurement of a Boolean
function of $\hat{B}$), performed alone, must induce a projection of the
output state (\ref{output}) on a state of the general form%
\begin{equation*}
\left\vert \chi \right\rangle =\frac{1}{\sqrt{c^{\prime }}}\dsum\limits_{%
\mathbf{b~\in ~}\sigma _{B}^{\prime }}\func{e}^{i\varphi _{\mathbf{b}%
}}\left\vert \mathbf{b}\right\rangle _{B}\left\vert s\left( \mathbf{b}%
\right) \right\rangle _{A}\left\vert \varphi \left( \mathbf{b}\right)
\right\rangle _{W},
\end{equation*}%
where $\sigma _{B}^{\prime }$\ is a subset of $\sigma _{B}$ of cardinality $%
c^{\prime }$. Alice's advanced knowledge is obtained by advancing by $%
U^{\dag }$ the two ends of this projection at the input of the quantum
algorithm, at time $t_{1}$ immediately after the preparation of the problem
setting (which here is the outcome of Bob's measurement). Even without
knowing $U^{\dag }$, we know that, for equation (\ref{indi}), this projects
the input state (\ref{input}) on:%
\begin{equation}
\frac{1}{\sqrt{c^{\prime }}}\left( \dsum\limits_{\mathbf{b~\in ~}\sigma
_{B}^{\prime }}\func{e}^{i\varphi _{\mathbf{b}}}\left\vert \mathbf{b}%
\right\rangle _{B}\right) \left\vert 00...\right\rangle _{A}\left\vert \psi
\right\rangle _{W}.  \label{bo}
\end{equation}%
In particular, it projects the maximally mixed state of register $B$ (\ref%
{red}) on the state of lower entropy%
\begin{equation}
\frac{1}{\sqrt{c^{\prime }}}\dsum\limits_{\mathbf{b~\in ~}\sigma
_{B}^{\prime }}\func{e}^{i\varphi _{\mathbf{b}}}\left\vert \mathbf{b}%
\right\rangle _{B},  \label{ject}
\end{equation}%
which represents Alice's advanced knowledge -- Alice knows in advance that $%
\mathbf{b}\in \sigma _{B}^{\prime }$. For short we say that Alice's
measurement of $\hat{B}_{j}$ projects $\sigma _{B}$ on $\sigma _{B}^{\prime
} $.

Still for equation (\ref{indi}), the same projection can be obtained by
measuring $\hat{B}_{j}$\ in the input state. We also note that,
mathematically, nothing changes if we assume to start with that the two
complete measurements reduce to two partial measurements, of $\hat{B}_{i}$
and $\hat{B}_{j}$ , both performed in the input state. Conditions \textbf{I}
and \textbf{II }define the same pairs of partial observables $\hat{B}_{i}$
and $\hat{B}_{j}$ no matter whether $\hat{B}_{j}$ is measured in the input
or output state. In fact moving the measurement from the output to the input
state\ leaves all selections and reductions of the entropy of the solution
unaltered.

This latter way of assessing Alice's advanced knowledge highlights a
symmetry hidden in the former one. We are left with two partial measurements
of the content of register $B$ that satisfy conditions \textbf{I} and 
\textbf{II}, both performed in the input state. We can loose the memory of
which partial measurement is performed by Alice and which by Bob. Evidently,
either partial measurement can be the one performed by Alice. Therefore,
given a pair of partial measurements, of $\hat{B}_{i}$ and $\hat{B}_{j}$, in
the input state (\ref{input}) that satisfy conditions \textbf{I} and \textbf{%
II}, either partial measurement performed alone projects the maximally mixed
state of register $B$ (\ref{red}) on an instance of Alice's advanced
knowledge. By the way, in this sense we can say that, with quantum
retrocausality $\mathcal{R}=\frac{1}{2}$, Alice knows "half" of the problem
setting in advance.

It is important to note that register $W$, which we have considered for
generality and could be necessary to construct the quantum algorithm, is not
involved in the definition of the pairs $\hat{B}_{i}$ and $\hat{B}_{j}$. Let
us recall the conditions their measurements [which can be both performed in
the input state (\ref{input})] are submitted to: (i) together, they select a
value of $\mathbf{b}$, (ii) the information acquired by either measurement
is not acquired by the other, (iii) they satisfy equation (\ref{equg}), and
(iv) they satisfy requirement (\ref{no}). Conditions (i), (ii) and (iv)
only\ involve the input state of register $B$, namely state (\ref{red}).
Also condition (iii) does not involve register $W$, as the reductions of the
entropy of the solution $\Delta \mathcal{E}_{A}\left( \hat{B}_{i}\right) $
and $\Delta \mathcal{E}_{A}\left( \hat{B}_{j}\right) $ concern the trace of
the output state (\ref{output}) over registers $B$ and $W$.

Therefore, to the end of determining $\hat{B}_{i}$ and $\hat{B}_{j}$, we can
work with $\left\vert \func{in}\right\rangle _{BA}$ and $\left\vert \func{out%
}\right\rangle _{BA}$, the traces over $W$ of $\left\vert \func{in}%
\right\rangle _{BAW}$ and $\left\vert \func{out}\right\rangle _{BAW}$; it
suffices to drop $\left\vert \psi \right\rangle _{W}$\ and the $\left\vert
\varphi \left( \mathbf{b}\right) \right\rangle _{W}$. States $\left\vert 
\func{in}\right\rangle _{BA}$ and $\left\vert \func{out}\right\rangle _{BA}$%
, in turn, can be written solely on the basis of the pairs $\mathbf{b}$ and $%
s\left( \mathbf{b}\right) $, namely of the oracle problem.

Since the quantum algorithm can be seen as a sum over classical histories in
each of which Alice knows in advance one of the possible halves of the
problem setting and performs the function evaluations still necessary to
identify the solution, given an oracle problem, we can know the number of
function evaluations required to solve it with quantum retrocausality $R=%
\frac{1}{2}$.

\subsubsection{Example of application}

We apply the present procedure to Deutsch's problem. Of course we should
ignore Deutsch algorithm.

Given the problem, namely all the pairs $\mathbf{b}$ (ranging over $%
00,01,10,11$) and $s\left( \mathbf{b}\right) $ ($0$ if the function is
constant, $1$ if it is balanced), we write down $\left\vert \func{in}%
\right\rangle _{BA}$ and $\left\vert \func{out}\right\rangle _{BA}$ [of
course we obtain the traces over register $V$\ of states (\ref{ina}) and (%
\ref{outa})].\ We do not need to know $U$. It suffices to know that there
can be a unitary transformation between input and output that satisfies
equation (\ref{indi}). Under conditions \textbf{I} and \textbf{II}, $%
\left\vert \func{in}\right\rangle _{BA}$ and $\left\vert \func{out}%
\right\rangle _{BA}$ define the pairs of partial observables $\hat{B}_{i}$
and $\hat{B}_{j}$ we are looking for -- it is easier to think they are both
measured in $\left\vert \func{in}\right\rangle _{BA}$. It is not a
constructive definition, however finding the pairs in question will be easy
in all the cases examined in this work. In the case of Deutsch's problem
they are any two of the three partial observables: $\hat{B}_{0}$, $\hat{B}%
_{1}$, and $\hat{B}_{X}\equiv \limfunc{XOR}\left( \hat{B}_{0},\hat{B}%
_{1}\right) $. These partial observables are Boolean functions of $\hat{B}$\
and the measurements of any two of them satisfy conditions \textbf{I} and 
\textbf{II} with $\Delta \mathcal{E}_{A}\left( \hat{B}_{i}\right) =\Delta 
\mathcal{E}_{A}\left( \hat{B}_{j}\right) =1/2$ bit.

Given a problem setting, say $\mathbf{b}=01$, either partial observable, $%
\hat{B}_{i}$ or $\hat{B}_{j}$, corresponds to an instance of Alice's
advanced knowledge as follows. We should assume that its measurement selects
the eigenvalue that matches with the problem setting. With problem setting $%
\mathbf{b\equiv ~}b_{0}b_{1}=01$, this implies that the measurement of $\hat{%
B}_{0}$ selects $b_{0}=0$, that of $\hat{B}_{1}$ selects $b_{1}=1$, and that
of $\hat{B}_{X}$ selects $\limfunc{XOR}\left( b_{0},b_{1}\right) =1$. The
corresponding projections of $\sigma _{B}$\ are respectively on $\left\{
00,01\right\} _{B}$, $\left\{ 01,11\right\} _{B}$, and $\left\{
01,10\right\} _{B}$. Thus the instances of Alice's advanced knowledge are $%
\mathbf{b}\in \left\{ 01,00\right\} _{B}$, $\mathbf{b}\in \left\{
01,11\right\} _{B}$, and $\mathbf{b}\in \left\{ 01,10\right\} _{B}$, as
obvious in hindsight. For any of these instances, Alice can solve the
problem with a single function evaluation.

We call the present procedure \textit{the advanced knowledge rule}. Given a
generic oracle problem, this rule defines the number of function evaluations
required to solve it with quantum retrocausality $\mathcal{R}=\frac{1}{2}$.
The importance of this rule depends on the confidence that can be placed in
the assumption that retrocausality $R=\frac{1}{2}$ is always attainable.
This is the case in all the quantum algorithms examined in the present work.
Whether it is the case in general should be the object of further work, the
present one is an exploration.

\section{Grover Algorithm}

Bob hides a ball in one of $N$\ drawers (ie, he marks an item in an
unstructured database of size $N$). Alice is to locate it by opening
drawers. In the classical case, to be a-priori certain of locating the ball,
Alice should plan to open $\limfunc{O}\left( N\right) $ drawers, in the case
of Grover $\left[ 30\right] $ quantum search algorithm $\limfunc{O}\left( 
\sqrt{N}\right) $. \ 

The problem, an oracle one, is formalized as follows. Let $\mathbf{b}$ and $%
\mathbf{a}$, belonging to $\left\{ 0,1\right\} ^{n}$, with $2^{n}=N$, be
respectively the number of the drawer with the ball and that of the drawer
that Alice wants to open. Checking whether the ball is in drawer $\mathbf{a}$
amounts to evaluating the function $f_{\mathbf{b}}\left( \mathbf{a}\right)
:\left\{ 0,1\right\} ^{n}\rightarrow \left\{ 0,1\right\} $, which is $1$ if $%
\mathbf{a=b}$ and $0$ otherwise.

Bob chooses one of the functions $f_{\mathbf{b}}\left( \mathbf{a}\right) $
(ie a value of $\mathbf{b}$) and gives Alice the black box that computes it.
Alice is to find the value of $\mathbf{b}$ chosen by Bob by performing
function evaluations for appropriate values of $\mathbf{a}$.

We will distinguish between $n=2$ and $n>2$. The speedup of Grover's
algorithm with $n=2$ is explained by $R=\frac{1}{2}$. When $n$ goes past $2$%
, $R$ slightly goes above $\frac{1}{2}$, to go back to $\frac{1}{2}$ for $%
n\rightarrow \infty $.

\subsection{Grover algorithm with $n=2$}

\subsubsection{Time-symmetric representation to Alice}

The input and output states of the quantum algorithm to Alice are
respectively:%
\begin{equation}
U_{B}\left\vert \psi \right\rangle =\left\vert \psi \right\rangle =\frac{1}{2%
\sqrt{2}}\left( \func{e}^{i\varphi _{0}}\left\vert 00\right\rangle _{B}+%
\func{e}^{i\varphi _{1}}\left\vert 01\right\rangle _{B}+\func{e}^{i\varphi
_{2}}\left\vert 10\right\rangle _{B}+\func{e}^{i\varphi _{3}}\left\vert
11\right\rangle _{B}\right) \left\vert 00\right\rangle _{A}\left( \left\vert
0\right\rangle _{V}-\left\vert 1\right\rangle _{V}\right) ,  \label{ing}
\end{equation}%
\begin{eqnarray}
\Im _{A}U_{f}H_{A}U_{B}\left\vert \psi \right\rangle &=&\frac{1}{2\sqrt{2}}%
\left( \func{e}^{i\varphi _{0}}\left\vert 00\right\rangle _{B}\left\vert
00\right\rangle _{A}+\func{e}^{i\varphi _{1}}\left\vert 01\right\rangle
_{B}\left\vert 01\right\rangle _{A}+\func{e}^{i\varphi _{2}}\left\vert
10\right\rangle _{B}\left\vert 10\right\rangle _{A}+\func{e}^{i\varphi
_{3}}\left\vert 11\right\rangle _{B}\left\vert 11\right\rangle _{A}\right) 
\notag \\
&&\left( \left\vert 0\right\rangle _{V}-\left\vert 1\right\rangle
_{V}\right) .  \label{outg}
\end{eqnarray}

The function of registers $B$, $A$, and $V$ is as in Deutsch algorithm.$\
U_{B}$ unitarily transforms the random outcome of Bob's measurement into the
desired problem setting, $H_{A}$ is the Hadamard transform on register $A$, $%
U_{f}$ is function evaluation, and $\Im _{A}$ -- a unitary transformation on
register $A$ -- is the so called \textit{inversion about the mean}. Note
that we could write the input and output states of registers $B$ and $A$
only on the basis of the pairs $\mathbf{b}$ and $s\left( \mathbf{b}\right) $
and without knowing Grover algorithm. The state of register $V$ is not
relevant for the determination of Alice's advanced knowledge.

Measuring $\hat{A}$\ in the output\ state (\ref{outg}) yields the number of
the drawer with \ the ball chosen by Bob.

\subsubsection{Quantum feedback}

We apply the advanced knowledge rule to Grover's problem with $n=2$. This
yields the number of function evaluations required to solve the problem with
quantum retrocausality $R=\frac{1}{2}$. The pairs of partial observables are
the same as in Deutsch algorithm: all the pairs among $\hat{B}_{0}$, $\hat{B}%
_{1}$, and $\hat{B}_{X}$. One can see that they satisfy conditions \textbf{I}
and \textbf{II} with $\Delta \mathcal{E}_{A}\left( \hat{B}_{i}\right)
=\Delta \mathcal{E}_{A}\left( \hat{B}_{j}\right) =1$ bit.

Say that the problem setting chosen by Bob is $\mathbf{b}$ $=01$ -- ie Bob
hides the ball in drawer $01$. The instances of Alice's advanced knowledge
are: $\mathbf{b}\in \left\{ 01,00\right\} _{B}$, $\mathbf{b}\in \left\{
01,11\right\} _{B}$, and $\mathbf{b}\in \left\{ 01,10\right\} _{B}$. In
other words, Alice knows in advance that the ball is in one of a pair
drawers (one of which with the ball in it). This allows her to locate the
ball by opening either drawer (ie by performing just one function
evaluation).

All the above could be derived solely from $\left\vert \func{in}%
\right\rangle _{BA}$ and $\left\vert \func{out}\right\rangle _{BA}$, the
traces over register $V$\ of states (\ref{ing}) and (\ref{outg}), which can
be written solely on the basis of the pairs $\mathbf{b}$ and $s\left( 
\mathbf{b}\right) $. One does not need to know Grover algorithm. However, it
is of course in agreement with the $n=2$ instance of Grover algorithm. This
means that the speedup of this instance is explained by quantum
retrocausality $R=\frac{1}{2}$.

We check that the present instance of Grover algorithm can be seen as a sum
over classical histories in each of which Alice knows in advance that the
ball is in a pair of drawers and locates it by opening either drawer. A
history is for example: 
\begin{equation}
\func{e}^{i\varphi _{1}}\left\vert 01\right\rangle _{B}\left\vert
00\right\rangle _{A}\left\vert 0\right\rangle _{V}\overset{H_{A}}{%
\rightarrow }\func{e}^{i\varphi _{1}}\left\vert 01\right\rangle
_{B}\left\vert 11\right\rangle _{A}\left\vert 0\right\rangle _{V}\overset{%
U_{f}}{\rightarrow }\func{e}^{i\varphi _{1}}\left\vert 01\right\rangle
_{B}\left\vert 11\right\rangle _{A}\left\vert 0\right\rangle _{V}\overset{%
\Im _{A}}{\rightarrow }\func{e}^{i\varphi _{1}}\left\vert 01\right\rangle
_{B}\left\vert 01\right\rangle _{A}\left\vert 0\right\rangle _{V}.
\label{hig}
\end{equation}%
The problem setting is $\mathbf{b}=01$. Alice performs function evaluation
for $\mathbf{a}=11$ (second and third state). Therefore we must assume that
Alice's advanced knowledge is $\mathbf{b}\in \left\{ 01,11\right\} _{B}$.
Since the output of function evaluation is zero (the content of register $V$%
\ remains unaltered), she finds that the number of the drawer with the ball
must be $\mathbf{b}=01$.

\subsection{Grover algorithm with $n>2$}

We should make a clarification to start with. With $n>2$, the original
Grover algorithm does not provide the solution of the problem with absolute
certainty. For this, one has to resort to the revisitation of Grover
algorithm made by Long $\left[ 31\right] $ -- see also $\left[ 32\right] $.
Long's algorithm can be tuned to provide the solution of Grover's problem
with certainty with any number of function evaluations provided it is above
the minimum number required by the optimal quantum algorithm, which is $K=%
\frac{\pi }{4\arcsin 2^{-n/2}}\approx \frac{\pi }{4}2^{n/2}$. Incidentally,
this is also the number required by Grover algorithm, which however does not
provide the solution with certainty when $n>2$.

With $\mathcal{R}=\frac{1}{2}$, the number of function evaluations required
by the present retrocausality model would be $2^{n/2}-1\approx 2^{n/2}$. In
fact, Alice knows in advance $\mathcal{R}n$ of the $n$ bits specify the
number of the drawer with the ball, thus (with $\mathcal{R}=\frac{1}{2}$) $%
n/2$ bits. This means that she must open in the worst case $2^{n/2}-1$
drawers (if all were empty, then she would know that the ball is in the only
drawer left).

We note anyhow that also the number of function evaluations foreseen by the
advanced knowledge rule, for $\mathcal{R}=\frac{1}{2}$, is that of an
existing quantum algorithm, which is in fact Long's algorithm tuned on $%
2^{n/2}-1$ function evaluations.

When $n$ goes past $2$, Alice's advanced knowledge should increase over the $%
n/2$ bits of the case $\mathcal{R}=\frac{1}{2}$, so that the problem can be
solved with $\approx \frac{\pi }{4}2^{n/2}$ function evaluations rather than 
$\approx 2^{n/2}$. This increase must be slight: an increase of just one bit
would halve the required number of function evaluations. Correspondingly, $%
\mathcal{R}$ should slightly go above $\frac{1}{2}$. It should also be noted
that, for $n\rightarrow \infty $, we have $\mathcal{R}=\frac{1}{2}$ again.

\section{Deutsch\&Jozsa algorithm}

Deutsch\&Jozsa $\left[ 33\right] $ algorithm is a generalization of the
seminal Deutsch algorithm that yields an exponential speedup. In the
respective problem, the set of functions is all the constant and \textit{%
balanced} functions (with the same number of zeroes and ones) $f_{\mathbf{b}%
}:\left\{ 0,1\right\} ^{n}\rightarrow \left\{ 0,1\right\} $. Array (\ref{dj}%
) gives the tables of four of the eight functions for $n=2$.%
\begin{equation}
\begin{tabular}{|l|l|l|l|l|l|}
\hline
$\mathbf{a}$ & ${\small \,f}_{0000}\left( \mathbf{a}\right) $ & ${\small f}%
_{1111}\left( \mathbf{a}\right) $ & ${\small f}_{0011}\left( \mathbf{a}%
\right) $ & ${\small f}_{1100}\left( \mathbf{a}\right) $ & ... \\ \hline
00 & 0 & 1 & 0 & 1 & ... \\ \hline
01 & 0 & 1 & 0 & 1 & ... \\ \hline
10 & 0 & 1 & 1 & 0 & ... \\ \hline
11 & 0 & 1 & 1 & 0 & ... \\ \hline
\end{tabular}
\label{dj}
\end{equation}

The bit string $\mathbf{b}\equiv b_{0}b_{1}...b_{2^{n}-1}$ is both the
suffix and the table of the function $f_{\mathbf{b}}\left( \mathbf{a}\right) 
$ -- the sequence of function values for increasing values of the argument.
Alice is to find whether the function chosen by Bob is constant or balanced
by computing $f_{\mathbf{b}}\left( \mathbf{a}\right) $ for appropriate
values of $\mathbf{a}$. Classically, this requires in the worst case a
number of function evaluations exponential in $n$. It requires just one
function evaluation in the quantum case.

\subsection{Time-symmetric representation to Alice}

The input and output states of the quantum algorithm to Alice are
respectively:%
\begin{equation*}
U_{B}\left\vert \psi \right\rangle =\left\vert \psi \right\rangle =\frac{1}{4%
}\left( \func{e}^{i\varphi _{0}}\left\vert 0000\right\rangle _{B}+\func{e}%
^{i\varphi _{1}}\left\vert 1111\right\rangle _{B}+\func{e}^{i\varphi
_{2}}\left\vert 0011\right\rangle _{B}+\func{e}^{i\varphi _{3}}\left\vert
1100\right\rangle _{B}+...\right) \left\vert 00\right\rangle _{A}\left(
\left\vert 0\right\rangle _{V}-\left\vert 1\right\rangle _{V}\right) ,
\end{equation*}%
\begin{eqnarray}
H_{A}U_{f}H_{A}U_{B}\left\vert \psi \right\rangle &=&\frac{1}{4}\left[
\left( \func{e}^{i\varphi _{0}}\left\vert 0000\right\rangle _{B}-\func{e}%
^{i\varphi _{1}}\left\vert 1111\right\rangle _{B}\right) \left\vert
00\right\rangle _{A}+\left( \func{e}^{i\varphi _{2}}\left\vert
0011\right\rangle _{B}-\func{e}^{i\varphi _{3}}\left\vert 1100\right\rangle
_{B}\right) \left\vert 10\right\rangle _{A}+...\right]  \notag \\
&&\left( \left\vert 0\right\rangle _{V}-\left\vert 1\right\rangle
_{V}\right) .  \label{tredj}
\end{eqnarray}%
Registers $B$, $A$, and $V$ and the unitary transformation $U_{B}$\ have the
same function as in the previous quantum algorithms. $H_{A}$ is the Hadamard
transform on register $A$ and $U_{f}$ is function evaluation. Note that we
could have written the input and output states of registers $B$ and $A$ only
on the basis of the pairs $\mathbf{b}$ and $s\left( \mathbf{b}\right) $.

Measuring $\hat{A}$\ in the output\ state (\ref{tredj}) says that the
function is constant if the measurement outcome is all zeros, balanced
otherwise.

\subsection{Quantum feedback}

We apply the advanced knowledge rule to Deutsch\&Jozsa's problem. Given the
problem setting of a balanced function, there is only one pair of partial
measurements of the content of register $B$ compatible with conditions 
\textbf{I} and \textbf{II}. With problem setting, say, $\mathbf{b}$ $=0011$, 
$\hat{B}_{i}$ must be the content of the left half of register $B$ and $\hat{%
B}_{j}$ that of the right half. The measurement of $\hat{B}_{i}$ yields all
zeros, that of $\hat{B}_{j}$ all ones.

In fact, a partial measurement yielding both zeroes and ones would violate
condition (\ref{no}): it would provide enough information to identify the
solution -- the fact that $f_{\mathbf{b}}$ is balanced. Given that either
partial measurement must yield all zeroes or all ones, it must concern the
content of half register. Otherwise either equation (\ref{equg}) would be
violated or the problem setting would not be completely determined, as
readily checked.\ 

One can see that, with $\mathbf{b}$ $=0011$, the measurement of $\hat{B}_{i}$%
, performed alone, projects $\sigma _{B}$ on the subset $\left\{
0011,0000\right\} _{B}$, that of $\hat{B}_{j}$ on$\ \left\{
0011,1111\right\} _{B}$. Either subset represents the part of the problem
setting that Alice knows in advance.\ Equation (\ref{equg}) is satisfied
with $\Delta \mathcal{E}_{A}\left( \hat{B}_{i}\right) =\Delta \mathcal{E}%
_{A}\left( \hat{B}_{j}\right) =1$ bit.

The case of the problem setting of a constant function is analogous. The
only difference is that there are more pairs of partial measurements that
satisfy the above said conditions. Say that the problem setting is $\mathbf{b%
}$ $=0000$. The measurements of the content of the left and right half of
register $B$ (each performed alone) projects $\sigma _{B}$ on respectively $%
\left\{ 0000,0011\right\} _{B}$ and $\left\{ 0000,1100\right\} _{B}$, the
measurements of the content of even and odd cells (say the leftmost one is
odd) on respectively $\left\{ 0000,0101\right\} _{B}$ and $\left\{
0000,1010\right\} _{B}$, etc.

There is a shortcut to finding the subsets in question. Here the problem
setting -- the bit string $\mathbf{b}$ -- is the table of the function
chosen by Bob. For example $\mathbf{b}=0011$ is the table $f_{\mathbf{b}%
}\left( 00\right) =0,f_{\mathbf{b}}\left( 01\right) =0,f_{\mathbf{b}}\left(
10\right) =1,f_{\mathbf{b}}\left( 11\right) =1$. We call "good half table"
any half table in which all the values of the function are the same. One can
see that good half tables are in one-to-one correspondence with the subsets
of $\sigma _{B}$ in question. For example, the good half table $f_{\mathbf{b}%
}\left( 00\right) =0,f_{\mathbf{b}}\left( 01\right) =0$ corresponds to the
subset $\left\{ 0011,0000\right\} _{B}$, is the identical part of the two
bit-strings in it. Thus, given a problem setting, ie an entire table, either
good half table, or identically the corresponding subset of $\sigma _{B}$,
is a possible instance of Alice's advanced knowledge.

Because of the structure of tables, given the advanced knowledge of a good
half table, the entire table and thus the solution can be identified by
performing just one function evaluation for any value of the argument $%
\mathbf{a}$ outside the half table.

Summing up, the advanced knowledge rule says that Deutsch\&Jozsa's problem
can be solved with just one function evaluation. This is in agreement with
Deutsch\&Jozsa algorithm, what also means that the speedup of this algorithm
is explained by quantum retrocausality $\mathcal{R}=\frac{1}{2}$.

We check that the present instance of Deutsch\&Jozsa algorithm can be seen
as a sum over classical histories in each of which Alice knows in advance
that Bob has chosen one of a pair of functions and discriminates between the
two with just one function evaluation. A history is for example: $\func{e}%
^{i\varphi _{2}}\left\vert 0011\right\rangle _{B}\left\vert 00\right\rangle
_{A}\left\vert 0\right\rangle _{V}\overset{H_{A}}{\rightarrow }\func{e}%
^{i\varphi _{2}}\left\vert 0011\right\rangle _{B}\left\vert 10\right\rangle
_{A}\left\vert 0\right\rangle _{V}\overset{U_{f}}{\rightarrow }\func{e}%
^{i\varphi _{2}}\left\vert 0011\right\rangle _{B}\left\vert 10\right\rangle
_{A}\left\vert 1\right\rangle _{V}\overset{H_{A}}{\rightarrow }\func{e}%
^{i\varphi _{2}}\left\vert 0011\right\rangle _{B}\left\vert 10\right\rangle
_{A}\left\vert 1\right\rangle _{V}$. Since the problem setting is $\mathbf{b}%
=0011$ and Alice performs function evaluation for $\mathbf{a}=10$, her
advanced knowledge must be $\mathbf{b}\in \left\{ 0011,0000\right\} _{B}$;
if it were $\mathbf{b}\in \left\{ 0011,1111\right\} _{B}$, she would have
performed function evaluation for either $\mathbf{a}=00$ or $\mathbf{a}=01$.
The result of function evaluation, $f_{\mathbf{b}}\left( 10\right) =1$,
tells that the function chosen by Bob is ${\small f}_{0011}\left( \mathbf{a}%
\right) $, hence that it is balanced.

One can see that the present analysis, like the notion of good half table,
holds unaltered for $n>2$.

\section{Simon and hidden subgroup algorithms}

In Simon's $\left[ 34\right] $ problem, the set of functions is all the $f_{%
\mathbf{b}}:\left\{ 0,1\right\} ^{n}\rightarrow \left\{ 0,1\right\} ^{n-1}$
such that $f_{\mathbf{b}}\left( \mathbf{a}\right) =f_{\mathbf{b}}\left( 
\mathbf{c}\right) $ if and only if $\mathbf{a}=\mathbf{c}$\ or $\mathbf{a}=%
\mathbf{c}\oplus \mathbf{h}\left( \mathbf{b}\right) $; $\oplus $\ denotes
bitwise modulo 2 addition. The bit string $\mathbf{h}\left( \mathbf{b}%
\right) $, depending on $\mathbf{b}$, is a sort of period of the function.

Array (\ref{periodic}) gives the tables of four of the six functions for $%
n=2 $. The bit string $\mathbf{b}$ is both the suffix and the table of the
function. We note that each value of the function appears exactly twice in
each table; thus 50\% of the rows plus one always identify $\mathbf{h}\left( 
\mathbf{b}\right) $.%
\begin{equation}
\begin{tabular}{|l|l|l|l|l|l|}
\hline
& $\mathbf{h}\left( 0011\right) =01$ & $\mathbf{h}\left( 1100\right) =01$ & $%
\mathbf{h}\left( 0101\right) =10$ & $\mathbf{h}\left( 1010\right) =10$ & ...
\\ \hline
$\mathbf{a}$ & ${\small f}_{0011}\left( \mathbf{a}\right) $ & ${\small f}%
_{1100}\left( \mathbf{a}\right) $ & ${\small f}_{0101}\left( \mathbf{a}%
\right) $ & ${\small f}_{1010}\left( \mathbf{a}\right) $ & ... \\ \hline
00 & 0 & 1 & 0 & 1 & ... \\ \hline
01 & 0 & 1 & 1 & 0 & ... \\ \hline
10 & 1 & 0 & 0 & 1 & ... \\ \hline
11 & 1 & 0 & 1 & 0 & ... \\ \hline
\end{tabular}
\label{periodic}
\end{equation}

Bob chooses one of these functions. Alice is to find the value of $\mathbf{h}%
\left( \mathbf{b}\right) $ by performing function evaluation\ for
appropriate values of $\mathbf{a}$.

In present knowledge, a classical algorithm requires a number of function
evaluations exponential in $n$. The quantum part of Simon algorithm solves
with just one function evaluation the hard part of this problem, namely
finding a string $\mathbf{s}_{j}\left( \mathbf{b}\right) $ \textit{orthogonal%
} $\left[ 34\right] $ to $\mathbf{h}\left( \mathbf{b}\right) $. There are $%
2^{n-1}$ such strings. Running the quantum part yields one of these strings
at random. The quantum part is iterated until finding $n-1$ different
strings. This allows Alice to find $\mathbf{h}\left( \mathbf{b}\right) $ by
solving a system of modulo 2 linear equations. Thus, on average, finding $%
\mathbf{h}\left( \mathbf{b}\right) $ requires $\limfunc{O}\left( n\right) $
iterations of the quantum part -- in particular $\limfunc{O}\left( n\right) $
function evaluations. Moreover, if we put an upper bound to the number of
iterations, a-priori there is always a non-zero probability of not finding $%
n-1$ different strings.

We apply the advanced knowledge rule directly to the complete Simon's
problem of finding $\mathbf{h}\left( \mathbf{b}\right) $ through function
evaluations. This is not the problem solved by the quantum part of Simon
algorithm, which is finding at random one of the $\mathbf{s}_{j}\left( 
\mathbf{b}\right) $ orthogonal to $\mathbf{h}\left( \mathbf{b}\right) $. The
value of $\mathcal{R}$\ that explains the speedup of the quantum part of
Simon algorithm will be a by-product of applying the advanced knowledge rule
to Simon's problem.

\subsection{Time-symmetric representation to Alice}

Knowing all the pairs $\mathbf{b}$, $\mathbf{h}\left( \mathbf{b}\right) $ --
from array (\ref{periodic}) -- we can write $\left\vert \func{in}%
\right\rangle _{BA}$ and $\left\vert \func{out}\right\rangle _{BA}$:

\begin{equation*}
\left\vert \func{in}\right\rangle _{BA}=\frac{1}{\sqrt{6}}\left( \func{e}%
^{i\varphi _{0}}\left\vert 0011\right\rangle _{B}+\func{e}^{i\varphi
_{1}}\left\vert 1100\right\rangle _{B}+\func{e}^{i\varphi _{2}}\left\vert
0101\right\rangle _{B}+\func{e}^{i\varphi _{3}}\left\vert 1010\right\rangle
_{B}+...\right) \left\vert 00\right\rangle _{A},
\end{equation*}%
\begin{equation*}
\left\vert \func{out}\right\rangle _{BA}=\frac{1}{\sqrt{6}}\left[ \left( 
\func{e}^{i\varphi _{0}}\left\vert 0011\right\rangle _{B}+\func{e}^{i\varphi
_{1}}\left\vert 1100\right\rangle _{B}\right) \left\vert 01\right\rangle
_{A}+\left( \func{e}^{i\varphi _{2}}\left\vert 0101\right\rangle _{B}+\func{e%
}^{i\varphi _{3}}\left\vert 1010\right\rangle _{B}\right) \left\vert
10\right\rangle _{A}+...\right] .
\end{equation*}%
We must assume that\ there can be a unitary transformation between the
un-traced states $\left\vert \func{in}\right\rangle _{BAW}$ and $\left\vert 
\func{out}\right\rangle _{BAW}$.

\subsection{Quantum feedback}

The analysis is similar to that of Deutsch\&Jozsa algorithm. This time a
good half table should not contain a same value of the function twice, what
would provide enough information to identify the solution of the problem [ie
the \textit{period} $\mathbf{h}\left( \mathbf{b}\right) $], thus violating
condition (\ref{no}) of the advanced knowledge rule.

With $\mathbf{b}=0011$, namely $f_{\mathbf{b}}\left( 00\right) =0,~f_{%
\mathbf{b}}\left( 10\right) =1,f_{\mathbf{b}}\left( 01\right) =0,~f_{\mathbf{%
b}}\left( 11\right) =1$, one way of sharing the table into two good halves
is: $f_{\mathbf{b}}\left( 00\right) =0,~f_{\mathbf{b}}\left( 10\right) =1$
and $f_{\mathbf{b}}\left( 01\right) =0,~f_{\mathbf{b}}\left( 11\right) =1$.
The corresponding subsets of $\sigma _{B}$\ are respectively $\left\{
0011,0110\right\} _{B}$ and $\left\{ 0011,1001\right\} _{B}$; one can check
that each half table is the identical part of the two bit-strings in the
corresponding subset of $\sigma _{B}$. Either good half table or identically
either subset is a possible instance of Alice's advanced knowledge. Equation
(\ref{equg}) is satisfied with $\Delta \mathcal{E}_{A}\left( \hat{B}%
_{i}\right) =\Delta \mathcal{E}_{A}\left( \hat{B}_{j}\right) =0.585$ bit
(entropy reduction from $-\log _{2}\frac{1}{3}$ bit to $1$ bit).

We note parenthetically that sharing each table into two halves is
accidental to Deutsch\&Jozsa's and Simon algorithms. In the quantum part of
Shor's $\left[ 35\right] $ factorization algorithm (finding the period of a
periodic function), taking two shares of the table that do not contain a
same value of the function twice implies that each share is less than half
table if the domain of the function spans more than two periods.

Given the advanced knowledge of a good half table, the entire table and then 
$\mathbf{h}\left( \mathbf{b}\right) $ can always be identified by performing
just one function evaluation for any value of the argument $\mathbf{a}$
outside the half table. Thus, the advanced knowledge rule says that, with $%
\mathcal{R}=\frac{1}{2}$, Simon's problem is solved with just one function
evaluation. Under the assumption that $\mathcal{R}=\frac{1}{2}$ is always
attainable, Simon algorithm, which requires $\limfunc{O}\left( n\right) $
function evaluations, would be suboptimal.

The above also shows that the speedup of the quantum part of Simon algorithm
is explained by $\mathcal{R}=\frac{1}{2}$. In fact, once known $\mathbf{h}%
\left( \mathbf{b}\right) $ -- with just one function evaluation in the case
of quantum retrocausality $\mathcal{R}=\frac{1}{2}$ -- generating at random
the $\mathbf{s}_{j}\left( \mathbf{b}\right) $'s requires no further function
evaluations.

We give the simplest instance, $n=2$, of the quantum algorithm that finds $%
\mathbf{h}\left( \mathbf{b}\right) $ with just one function evaluation.
Register $W$ reduces to the usual register $V$ that contains the result of
function evaluation modulo 2 added to its previous content. The input and
output states of $V$ are both $\frac{1}{\sqrt{2}}\left( \left\vert
0\right\rangle _{V}-\left\vert 1\right\rangle _{V}\right) $.\ We have $U=%
\mathcal{P}_{A}H_{A}U_{f}H_{A}$, where $H_{A}$ is Hadamard on register $A$, $%
U_{f}$ function evaluation, $\mathcal{P}_{A}$ the permutation of the basis
vectors $\left\vert 01\right\rangle _{A}$ and $\left\vert 10\right\rangle
_{A}$. Checking whether there is the similar algorithm for $n>2$ should be
the object of further work.

The sum over histories representation can be developed as in Deutsch\&Jozsa
algorithm. If, for example, Alice's advanced knowledge is $\mathbf{b}\in
\left\{ 0011,0110\right\} _{B}$, she can identify the value of $\mathbf{h}%
\left( \mathbf{b}\right) $ by performing a single function evaluation for
either $\mathbf{a}=01$ or $\mathbf{a}=11$ -- see array (\ref{periodic}) --
etc.

The fact that Alice knows in advance a good half table, and can thus
identify the entire table and hence the solution with just one function
evaluation, clearly holds unaltered for $n>2$. It should also apply to the
generalized Simon's problem and to the Abelian hidden subgroup problem. In
fact the corresponding algorithms are essentially Simon algorithm. In the
hidden subgroup problem, the set of functions $f_{\mathbf{b}}:G\rightarrow W$
map a group $G$ to some finite set $W$\ with the property that there exists
some subgroup $S\leq G$ such that for any $\mathbf{a},\mathbf{c}\in G$, $f_{%
\mathbf{b}}\left( \mathbf{a}\right) =f_{\mathbf{b}}\left( \mathbf{c}\right) $
if and only if $\mathbf{a}+S=\mathbf{c}+S$. The problem is to find the
hidden subgroup $S$ by computing $f_{\mathbf{b}}\left( \mathbf{a}\right) $
for the appropriate values of $\mathbf{a}$. Now, a large variety of problems
solvable with a quantum speedup can be re-formulated in terms of the hidden
subgroup problem $\left[ 11\right] $. Among these we find: the seminal
Deutsch's problem, finding orders, finding the period of a function (thus
the problem solved by the quantum part of Shor's\ factorization algorithm),
discrete logarithms in any group, hidden linear functions, self shift
equivalent polynomials, Abelian stabilizer problem, graph automorphism
problem $\left[ 36\right] $.

\section{Conclusion}

We have extended the representation of the quantum algorithm to the process
of setting the problem. The initial measurement, in a state where the
problem setting is completely undetermined, selects a setting at random, a
unitary evolution transforms it into the desired setting, a further unitary
evolution solves the problem, and the final measurement reads the solution.

This extended representation would tell Alice, the problem solver, the
setting of the problem before she begins her search for the solution. To
Alice, this setting must be hidden inside the black box. To physically
represent this concealment, we resorted to relational quantum mechanics. In
the representation relativized to Alice, the projection induced by the
initial measurement is retarded at the end of the unitary part of her
problem solving action. To Alice, the setting remains completely
undetermined throughout that part of her action.

In this time-symmetric representation of the quantum algorithm, the solution
of the problem is selected by either the initial Bob's measurement or the
final Alice's measurement. We assumed that the selection shares without
redundancies between the two measurements. This turned out to be equivalent
to sharing between initial and final measurement the selection of the random
outcome of the initial measurement. We have called $\mathcal{R}$ the
fraction of the information that specifies this random outcome whose
selection is ascribed to the final measurement. $\mathcal{R}$ is a measure
of retrocausality; $\mathcal{R}=0$ means that the random outcome of the
initial measurement is entirely selected by that same measurement, without
retrocausality. $\mathcal{R}=1$ means that it is entirely selected by the
final measurement, without time-forward causality. $\mathcal{R}=\frac{1}{2}$%
\ means that the selection equally shares between the two measurements.

The sharing in question is without consequences in the representation of the
quantum algorithm with respect to Bob and any external observer, where it
leaves the input state of the quantum algorithm unaltered. It projects the
input state relativized to Alice, one of maximal ignorance of the problem
setting, on a state of lower entropy where she knows the $\mathcal{R}$-the
part\ of the problem setting in advance, before performing any function
evaluation.

The quantum algorithm turns out to be a sum over classical histories in each
of which Alice knows in advance one of the possible $\mathcal{R}$-th parts
of the problem setting and performs the function evaluations logically
required to identify the solution. The number of function evaluations is
therefore that of a classical algorithm that benefits of the same advanced
knowledge.

Given an oracle problem and a value of $\mathcal{R}$, the present
explanation of the speedup provides the number of function evaluations
required to solve it quantumly. Conversely, given a known quantum algorithm,
it provides the value of $\mathcal{R}$\ that explains its speedup.

We have compared this explanation of the speedup with the major quantum
algorithms. $\mathcal{R}=\frac{1}{2}$ explains the speedup of the seminal
Deutsch algorithm, of Grover quantum search algorithm for database size $4$,
Deutsch\&Jozsa algorithm, and the algorithms of Simon and the Abelian hidden
subgroup. All these algorithms require a single function evaluation.

When database size goes past $4$, Grover algorithm requires more than one
function evaluation and $\mathcal{R}$ goes slightly above $\frac{1}{2}$,
going back to $\frac{1}{2}$ as database size goes to infinity.

In any way, the number of function evaluations foreseen by the present
explanation with $\mathcal{R}=\frac{1}{2}$ is always that of an existing
quantum algorithm and a good approximation of the number required by the
optimal one.

If the sample of quantum algorithms examined were representative enough,
namely if quantum retrocausality $\mathcal{R}=\frac{1}{2}$ were always
attainable in quantum problem solving, we would have a very powerful tool
for the study of quantum query complexity, a still open problem. This work
is an exploration. Wether $\mathcal{R}=\frac{1}{2}$ is always attainable,
what is the maximum value of $\mathcal{R}$ physically attainable are
questions that remain open.

$\mathcal{R}$, the fraction of the information that specifies the random
outcome of the initial measurement whose selection can mathematically be
ascribed to the final measurement, would seem to be a potentially
interesting retrocausality measure. Studying it from a foundational
standpoint might be rewarding. Because of the fundamental character of
quantum search in an unstructured database, one could conjecture that $%
\mathcal{R}=\frac{1}{2}$ is always attainable and that the maximum possible
value of $\mathcal{R}$ is attained in Grover algorithm. Whether this is so
and why should be the object of further work.

Another issue that might deserve further investigation is the possible
relation between $\mathcal{R}$\ and the information theoretic temporal Bell
inequalities.

From a practical standpoint, one should further study the trust that can be
placed in the validity of the $\mathcal{R}=\frac{1}{2}$ approximation by
checking the relation between speedup and $\mathcal{R}$ on larger classes of
known quantum algorithms. One could also investigate whether there is the
optimal quantum algorithm foreseen in Section 8 for Simon's and the Abelian
hidden subgroup problems.

\subsection*{Acknowledgments}

$\ \ $Thanks are due to David Finkelstein for useful discussions.

\subsection*{References}

$\left[ 1\right] $ Deutsch D. Quantum Theory, the Church Turing Principle
and the Universal Quantum Computer. Proceedings of the Royal Society of
London A 1985; 400: 97-117. doi:10.1098/rspa.1985.0070

$\left[ 2\right] $ Finkelstein D R. Space-time structure in high energy
interactions. In Gudehus T, Kaiser G, Perlmutter A editors, Fundamental
Interactions at High Energy. New York: Gordon \& Breach, 1969 pp. 324-338,
https://www.researchgate.net/publication/23919490\_Space-time\_structure\_in%
\_high\_energy\_interactions

$\left[ 3\right] $ Feynman R P. Simulating Physics with Computers.
International Journal of Theoretical Physics 1982; 21 (6--7): 467-488.
doi:10.1007/BF02650179,

https://www.cs.princeton.edu/courses/archive/fall05/frs119/papers/feynman82/feynman82.html

$\left[ 4\right] $ Bennett C H. The Thermodynamics of Computation -- a
Review. International Journal of Theoretical Physics 1982; 21, 905-940

$\left[ 5\right] $ Landauer R. Irreversibility and heat generation in the
computing process. IBM Journal of Research and Development 1961; 5 (3):
183--191, doi:10.1147/rd.53.0183, retrieved 2015-02-18,

http://www.pitt.edu/\symbol{126}jdnorton/lectures/Rotman\_Summer\_School%
\_2013/thermo\_computing\_docs/Landauer\_1961.pdf

$\left[ 6\right] $ Fredkin E, Toffoli T. Conservative Logic. International
Journal of Theoretical Physics 1982; 21: 219-253

$\left[ 7\right] $ Feynman, R P. Quantum mechanical computers. Foundations
of Physics 1986: Vol. 16, No. 6: 507-5031

$\left[ 8\right] $ Turin Elsag Bailey-ISI international worksohps on quantum
communication and computation. Years 1993, 1995, 1997 group pictures,

http://www.giuseppecastagnoli.com/images

$\left[ 9\right] $\ Rovelli C. Relational Quantum Mechanics. Int. Journal of
Theoretical Physics 1996; 35: 637-658. doi:10.1007/BF02302261

$\left[ 10\right] $ Mosca M, Ekert A. The Hidden Subgroup Problem and
Eigenvalue Estimation on a Quantum Computer. Proceedings QCQC '98, selected
papers from the First NASA International Conference on Quantum Computing and
Quantum Communications, Springer-Verlag London, UK, 1998, pp. 174-188.

$\left[ 11\right] $ Castagnoli G, Finkelstein DR. Theory of the quantum
speedup.\textit{\ }Proceedings of the Royal Society of London A 2001; 457:
1799-1807. doi:10.1098/rspa.2001.0797

$\left[ 12\right] $\ Castagnoli G. The quantum correlation between the
selection of the problem and that of the solution sheds light on the
mechanism of the quantum speed up. Physical Review A 2010; 82: 052334-052342.

$\left[ 13\right] $ Castagnoli G. Probing the mechanism of the quantum
speed-up by time-symmetric quantum mechanics. Proceedings of the 92nd\
Annual Meeting of the AAAS Pacific Division, Quantum Retrocausation: Theory
and Experiment, 2011.

$\left[ 14\right] $\ Aharonov Y, Bergman PG, Lebowitz JL. Time Symmetry in
the Quantum Process of Measurement. Physical Review B 1964; 134:\textbf{\ }%
1410-1416.

$\left[ 15\right] $ Aharonov Y, Albert D, Vaidman L. How the result of a
measurement of a component of the spin of a spin-1/2 particle can turn out
to be 100. Physical Review Letters 1988; 60 (14): 1351-1354.
http://dx.doi.org/10.1103/PhysRevLett.60.1351

$\left[ 16\right] $ Aharonov Y, Popescu S,\ Tollaksen J. A time-symmetric
formulation of quantum mechanics. Physics today 2010; November issue: 27-32.
doi:10.1063/1.3518209

$\left[ 17\right] $\ Dolev S, Elitzur AC. Non-sequential behavior of the
wave function. 2001; arXiv:quant-ph/0102109 v1

$\left[ 18\right] $ Morikoshi F. Problem-Solution Symmetry in Grover's
Quantum Search Algorithm. International Journal of Theoretical Physics 2011;
50: 1858-1867.

$\left[ 19\right] $ Morikoshi F. Information-theoretic temporal Bell
inequality and quantum computation. Physical Reviews A 2006; 73: 052308-
052312. doi:10.1007/s10773-011-0701-6

$\left[ 20\right] $ Jozsa R, Linden N. On the role of entanglement in
quantum computational speed-up. Proceedings of the Royal Society A 2002;
doi: 10.1098/rspa.2002.1097.

$\left[ 21\right] $ Ollivier H, Zurek WH. Quantum Discord: A Measure of the
Quantumness of Correlations. Physical Review Letters 2001; 88:
017901-017909. doi: http://dx.doi.org/10.1103/PhysRevLett.88.017901

$\left[ 22\right] $ Henderson L, Vedral V. Classical, quantum and total
correlations. Journal of Physics A 2001; 34: 6899- 6709

$\left[ 23\right] $ Gross D, Flammia S T, Eisert J. Most Quantum States Are
Too Entangled To Be Useful As Computational Resources. Physical Review
Letters 2009; 102. doi: http://dx.doi.org/10.1103/PhysRevLett.102.190501.

$\left[ 24\right] $ Cai Y, Le H N, Scarani V. State complexity and quantum
computation. Annalen der Physik 2015; x: Y. doi:10.1002/andp.201400199,
arxiv: quant-ph/1503.04017

$\left[ 25\right] $ Howard M, Wallman J, Veitch V, Emerson J. Contextuality
supplies the `magic' for quantum computation. Nature19 June 2014; 510,
351--355. doi:10.1038/nature13460

$\left[ 26\right] $ Bohm D, Pines DA. Collective Description of Electron
Interactions: III. Coulomb Interactions in a Degenerate Electron Gas.
Physical Review 1953; 92: 626-636.
doi:http://dx.doi.org/10.1103/PhysRev.92.609

$\left[ 27\right] $ Hawking S. On the Shoulders of Giants. Running Press,
Philadelphia-London 2003$.$

$\left[ 28\right] $ Finkelstein DR, private communication.

$\left[ 29\right] $ Feynman R, Hibbs AR. Quantum Mechanics And Path
Integrals. New York, McGraw-Hill 1965.

$\left[ 30\right] $ Grover LK. A fast quantum mechanical algorithm for
database search. Proceedings of the 28th Annual ACM Symposium on the Theory
of Computing. ACM press New York, 1996, pp. 212-219.

$\left[ 31\right] $ Long GL. Grover algorithm with zero theoretical failure
rate. Physical Review A 2001; 64: 022307-022314. doi:

http://dx.doi.org/10.1103/PhysRevA.64.022307

$\left[ 32\right] $ Toyama, FM, van Dijk W, Nogami Y. Quantum search with
certainty based on modified Grover algorithms: optimum choice of parameters.
Quantum Information Processing 2013; 12: 1897-1914.
doi:10.1007/s11128-012-0498-0

$\left[ 33\right] $\ Deutsch D, Jozsa R. Rapid Solution of Problems by
Quantum Computation. Proceedings of the Royal Society of London A 1992: 439:
553-558. doi:10.1098/rspa.1992.0167

$\left[ 34\right] $ Simon D. On the power of quantum computation.
Proceedings of the 35th Annual IEEE Symposium on the Foundations of Computer
Science, 1994, pp. 116-123.

$\left[ 35\right] $ Shor PW. Algorithms for quantum computation: discrete
logarithms and factoring. Proceedings of the 35th Annual IEEE Symposium on
the Foundations of Computer Science, 1994, pp. 124-134.

$\left[ 36\right] $ Kaye P, Laflamme R, Mosca M. An Introduction To Quantum
Computing. Oxford University Press, 2007; pp. 146-147.

\end{document}